\documentclass[11pt]{elsarticle}
\biboptions{round,authoryear,semicolon}
\renewcommand{\cite}{\citep} 

\usepackage{titlesec}
\titleformat*{\section}{\Large\bfseries}
\titleformat*{\subsection}{\large\bfseries}
\titleformat*{\subsubsection}{\normalsize\itshape}

\usepackage{amsmath}
\usepackage{amssymb}
\usepackage{amsthm}

\usepackage{graphicx}
\usepackage[labelfont=bf]{caption}
\usepackage{subcaption}

\usepackage{color}
\usepackage{soul}

\usepackage{diagbox}

\usepackage{float}
\floatstyle{boxed}
\newfloat{glosbox}{thp}{glosbox}
\floatname{glosbox}{Box}
\floatstyle{plain}

\usepackage{floatpag}
\floatpagestyle{empty}

\usepackage{tocloft}
\cftpagenumbersoff{figure}
\cftpagenumbersoff{table}
\usepackage{tikz}
\usetikzlibrary{arrows}

\makeatletter

\tikzset{mathterm/.style={draw=black,fill=white,rectangle,anchor=base}}
\tikzstyle{every picture}+=[remember picture]
\everymath{\displaystyle}

\newcommand\indicate[2][black]{%
   \tikz [baseline] \node [inner sep=0pt,anchor=base] (i#2) {\vphantom|};
   \@ifnextchar[{\@indicateopts{#1}{#2}}{\@indicatenoopts{#1}{#2}}}
\def\@indicatenoopts#1#2{%
   {\color{#1} \tikz[overlay] \path[line width=1pt,draw=#1,-stealth] (i#2) edge (#2);}}
\def\@indicateopts#1#2[#3]{%
   {\color{#1} \tikz[overlay] \path[line width=1pt,draw=#1,-stealth] (i#2) [#3] edge (#2);}}

\makeatother

\theoremstyle{definition}

\theoremstyle{remark}

\numberwithin{equation}{section}

\journal{Cell Systems}

\begin{document}

\begin{frontmatter}

\title{ 
Entropy-scaling search of massive biological data}

\author[mitmath,mitcsail]{Y. William Yu\corref{co}}
\author[mitmath,mitcsail]{Noah M. Daniels\corref{co}}
\author[mitcsail]{David Christian Danko}
\author[mitmath,mitcsail]{Bonnie Berger\corref{correspond}}
\ead{bab@mit.edu}
\cortext[co]{These authors contributed equally to this work.}
\cortext[correspond]{Corresponding author}
\address[mitmath]{Department of Mathematics, Massachusetts Institute of Technology, Cambridge, Massachusetts 02139}
\address[mitcsail]{Computer Science and AI Lab, Massachusetts Institute of Technology, Cambridge, Massachusetts 02139}






\begin{abstract}
    \begin{itemize}
        \item We describe entropy-scaling search for finding approximate matches in a database
        \item Search complexity is bounded in time and space by the entropy of the database
        \item We make tools that enable search of three largely intractable real-world databases
        \item The tools dramatically accelerate metagenomic, chemical, and protein structure search
    \end{itemize}
\noindent\unskip\textbf{eTOC Blurb}
\par\medskip\noindent\unskip\ignorespaces
We describe a general
framework for efficiently searching
massive datasets having certain
properties common in biology.
\end{abstract}

\end{frontmatter}

\section{Summary}
{ \bfseries
    Many datasets exhibit a well-defined structure that can be exploited to design faster search tools, but it is not always clear when such acceleration is possible.
    Here, we introduce a framework for similarity search based on characterizing a dataset's entropy and fractal dimension.
    We prove that searching scales in time with metric entropy (number of covering hyperspheres), if the fractal dimension of the dataset is low, and scales in space with the sum of metric entropy and information-theoretic entropy (randomness of the data).
    Using these ideas, we present accelerated versions of standard tools, with no loss in specificity and little loss in sensitivity, for use in three domains---high-throughput drug screening (Ammolite, 150$\times$ speedup), metagenomics (MICA, 3.5$\times$ speedup of DIAMOND [3,700$\times$ BLASTX]), and protein structure search (esFragBag, 10$\times$ speedup of FragBag).
    Our framework can be used to achieve ``compressive omics,'' and the general theory can be readily applied to data science problems outside of biology.

    Source code: \url{http://gems.csail.mit.edu}
}

\section{Introduction}
Throughout all areas of data science, researchers are confronted with increasingly large volumes of data. 
In many fields, this increase is exponential in nature, outpacing Moore's and Kryder's laws on the respective doublings of transistors on a chip and long-term data storage density \cite{kahn2011future}.
As such, the challenges posed by the massive influx of data cannot be solved by waiting for faster and larger capacity computers but, instead, require instead the development of data structures and representations that exploit the complex
structure of the dataset.

Here, we focus on similarity search, where the task at hand is to find all entries in some database that are ``similar,'' or approximate matches, to a query item.
Similarity search is a fundamental operation in data science and lies at the heart of many other problems, much like how sorting is a primitive operation in computer science.
Traditionally, approximate matching has been studied primarily in the context of strings under edit distance metrics (Box \ref{box:glos}) (e.g., for a spell-checker to suggest the most similar words to a misspelled word) \cite{ukkonen1985algorithms}.
Several approaches, such as the compressed suffix array and the FM-index~\cite{grossi2005compressed, ferragina2000opportunistic}, have been developed to accelerate approximate matching of strings.
However, it has been demonstrated that similarity search is also important in problem domains where biological data are not necessarily represented as strings, including computational screening of chemical graphs \cite{schaeffer2007graph} and searching protein structures \cite{budowski2010fragbag}.
Therefore, approaches that apply to more general conditions are needed.

As available data grow exponentially \cite{berger2013computational,yu2015quality} (e.g., genomic data in Figure S1), 
algorithms that scale linearly (Box \ref{box:glos}) with the amount of data no longer suffice.
The primary ways in which the literature addresses this problem---locality sensitive 
hashing \cite{indyk1998approximate}, vector approximation 
\cite{ferhatosmanoglu2000vector}, and space partitioning 
\cite{weber1998quantitative}---involve the construction of data structures that support more efficient search operations.
However, we note that, as biological data increase, not only does the redundancy present in the data also increase~\cite{loh2012compressive}, but also
internal structure (such as the fact that not all conceivable configurations, e.g. all possible protein sequences, actually exist) also becomes apparent.
Existing general-purpose methods do not explicitly exploit the particular 
properties of biological data to accelerate search (see the Theory section in the Supplemental Methods).

Previously, our group demonstrated how redundancy in genomic data could be used to accelerate local sequence alignment.
Using an approach that we called ``compresive genomics,'' we accelerated BLAST and BLAT \cite{kent2002blat} by taking advantage of high redundancy between related genomes using link pointers and edit scripts to a database of unique sequences \cite{loh2012compressive}.
We have used similar strategies to obtain equally encouraging results for local alignment in proteomics \cite{daniels2013compressive}.
Empirically, this compressive acceleration appears to scale almost linearly in the entropy of the database, often resulting in orders of magnitude better performance;
however, these previous studies neither proved complexity bounds nor established a theory to explain these empirical speedups.

Here, we generalize and formalize this approach by introducing 
a framework for similarity search of omics data.
We prove that search performance primarily depends on a measure of the novelty of new data, also known as entropy.
This framework, which we call entropy-scaling search, supports the creation of a data structure that provably scales linearly in both time and space with the entropy of the database, and thus sublinearly with the entire database.

We introduce two key concepts for characterizing a dataset: metric entropy and fractal dimension.
Intuitively, metric entropy measures how dissimilar the dataset is from itself, and fractal dimension measures how the number of spheres needed to cover all points in a database scales with the radii of those spheres.
Both are rigorously defined later, but note that metric entropy is not to be confused with the notion of a distance metric (Box \ref{box:glos}).
Using these two concepts, we show that, if similarity is defined by a metric-like distance function (e.g., edit or Hamming distance) and the database exhibits both low metric entropy and fractal dimension, the entropy-scaling search performs much better than naïve and even optimized methods.
Through three applications to large databases in chemogenomics, metagenomics, and protein structure search, we show that this framework allows for minimal (or even zero) loss in recall, coupled with zero loss in specificity.
The key benefit of formulating entropy-scaling search in terms of metric entropy and fractal dimension is that this allows us to provide mathematically rigorous guidance as to how to determine the efficacy of the approach for any dataset.

\begin{glosbox}[tbhp]
        \small
    \begin{itemize}
        \item Edit distance: the number of edits (character insertions, deletions, or substitutions) needed to turn one string into another.
        \item Scale, in time and space: the amount of time or space a task takes as a function of the amount of data on which it must operate. A task requiring time directly proportional to the size of the data is said to scale linearly; for example, searching a database takes twice as long if the database grows by a factor of two.
        \item Distance metric: a measure of distance that obeys several mathematical properties, including the triangle inequality.
        \item Covering spheres: a set of spheres around existing points so that every point is contained in at least one sphere and no sphere is empty.
        \item Metric entropy: a measure of how dissimilar a dataset is from itself. Defined as the number of covering spheres.
        \item Fractal dimension: a measure of how the number of points contained within a sphere scales with the radius of that sphere.
        \item Information-theoretic entropy: often used in data compression as shorthand for the number of bits needed to encode a database or a measure of the randomness of that database.
        \item Pattern matching: refers to searching for matches that might differ in specific ways from a query, such as wildcards or gaps, as opposed to searching for all database entries within a sphere of a specified radius as defined by an arbitrary distance function.
    \end{itemize}
    \caption{Definitions}
    \label{box:glos}
\end{glosbox}

\section{Results}

\subsection{Entropy-scaling similarity search}

\begin{figure}[p]
    \centering
    \centerline{\includegraphics[width=8in]{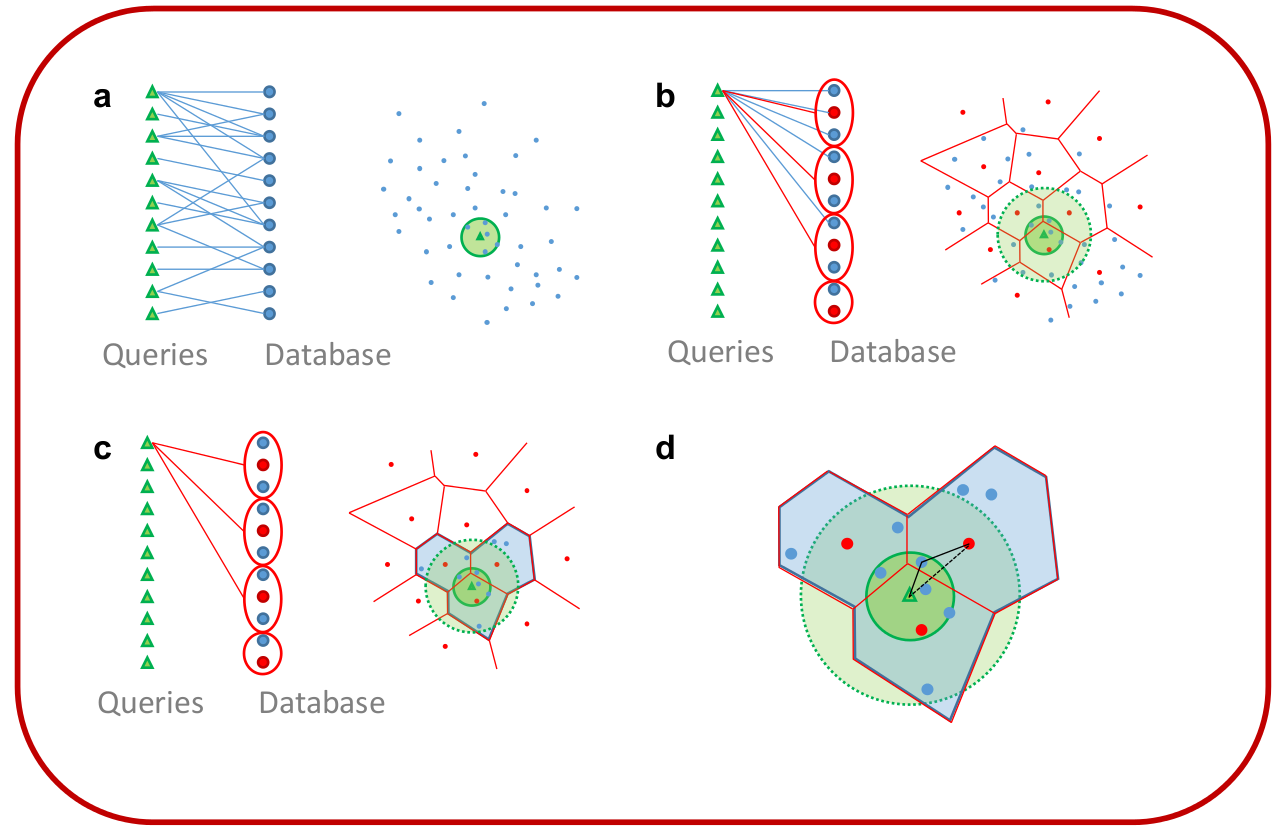}}
    \caption{ Entropy-scaling framework for similarity search. %
            (A-D) As shown, %
            (A) The na\"ive approach tests each query against each database entry to find entries within distance $r$ of the query (inside the small green disc). %
            (B) By selecting appropriate cluster centers with maximum radius $r_c$ to partition the database, we can (C) first do a coarse search to find all cluster centers within distance $r+r_c$ of a query (larger green disc), %
 and then the (D) triangle inequality guarantees that a fine search over all corresponding cluster entries (blue polygonal regions) will suffice.}
    \label{fig:dataStructure}
\end{figure}

The basic framework for the entropy-scaling search of a database involves four steps.
(1) Analyze the database to define a high-dimensional space and determine how to map database entries onto points in this space (this mapping may be one-to-one).
(2) Use this space and a measure of similarity between points to group entries in the database into clusters.
(3) To search for a particular query item, perform a coarse-grained search to identify the clusters that could possibly contain the query.
(4) Do a fine-grained search of the points contained within these clusters to find the closest matches to the query (Figure \ref{fig:dataStructure}).

Here, we provide conceptual motivation for this process.
In the following text, we consider entropy to be nearly synonymous with distance between points in a high-dimensional space; thus, with low entropy, newly added points do not tend to be far from all existing points. 
For genomic sequences, the distance function can be edit distance; for chemical graphs, it can be Tanimoto distance; and for general vectors, it can be Euclidean or cosine distance.
We are interested in the similarity search problem of finding all points in a set that are close to (i.e., similar to) the query point.

Let us first consider what it means for a large biological dataset, considered as points in a high-dimensional space, to be highly redundant.
Perhaps many of the points are exact duplicates; this easy scenario is trivially exploited by de-duplication and is already standard practice with datasets such as
the NCBI's non-redundant (NR) protein database \cite{pruitt2005ncbi}.
Maybe the points mostly live on a low-dimensional subspace; statistical tools such as Principal Component Analysis (PCA) exploit this property in data analysis.
Furthermore, if the dimension of the subspace is sufficiently low,
it can be divided into cells, allowing quick similarity searches by looking only at nearby cells \cite{weber1998quantitative}.
However, when the dimensionality of the subspace increases, cell search time 
grows exponentially; additionally, in sparse datasets, most of the cells will 
be empty, which wastes search time.

\begin{figure}[p]
    \vspace{-5em}
    \centering
    \centerline{\includegraphics[width=6in]{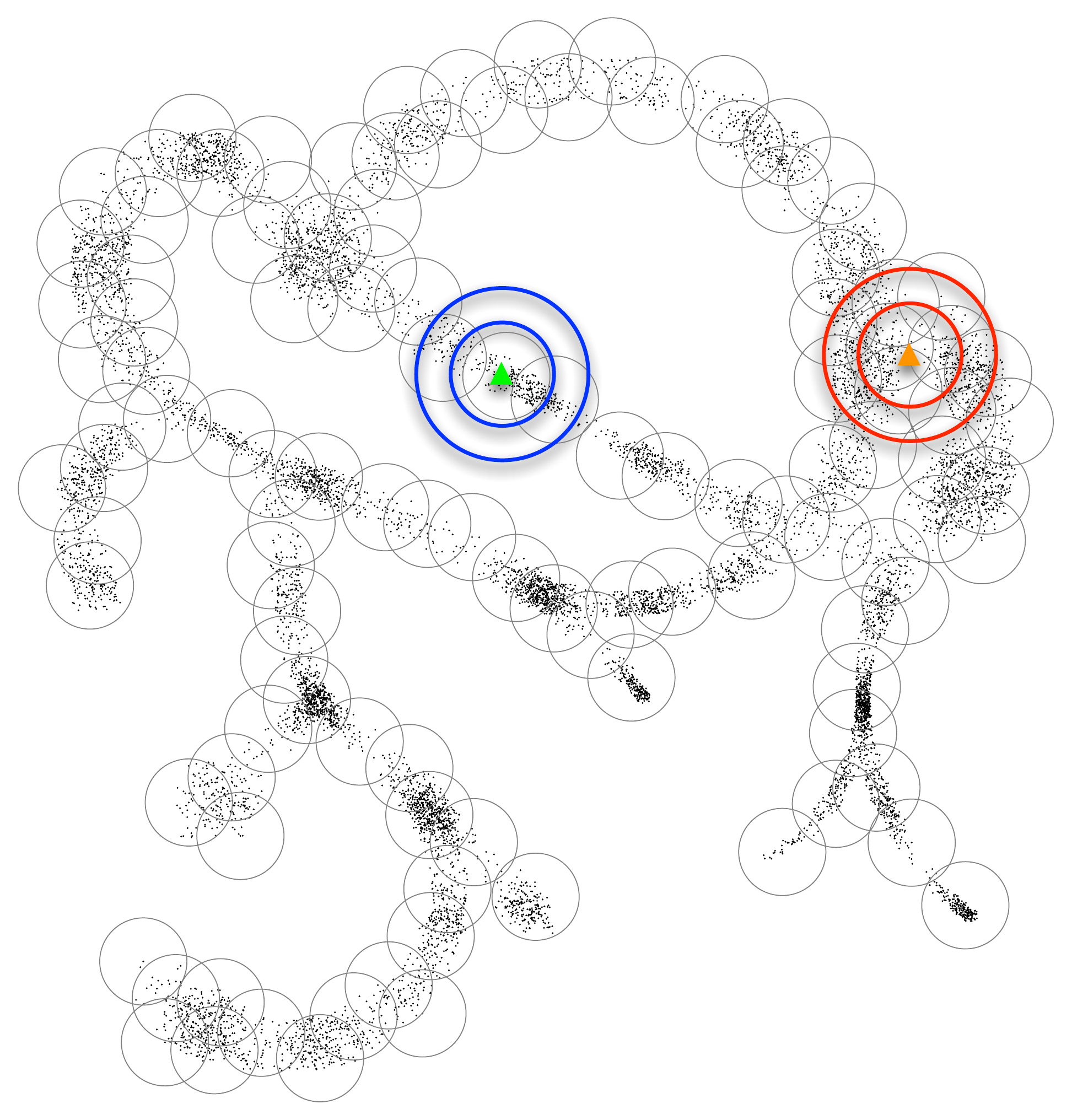}}
    \caption{Cartoon depiction of points in a high-dimensional space. %
        This cartoon depics points in an arbitrary high-dimensional space that live close to a one-dimensional tree-like structure, as might arise from genomes generated by mutation and selection along an evolutionary tree of life. %
Although high-dimensional at a fine scale, at the coarser scale of covering spheres, the data cloud looks nearly one-dimensional, which enables entropy-scaling of similarity search. The cluster center generation was performed using the same method we used for protein structure search. %
The blue circles around the green query point illustrate low fractal dimension: the larger-radius circle contains only linearly more points than the smaller one, rather than exponentially more. In contrast, the red circles around the orange query point illustrate higher local fractal dimension.}
    \label{fig:tree}
\end{figure}

More importantly, biological datasets generally do not live in low-dimensional subspaces.
Consider the instructive case of genomes along an evolutionary ``tree of life'' (Figure \ref{fig:tree}).
Such a tree has many branches (although admixture merges branches back together),
and looks nearly one-dimensional locally, but it is globally of higher dimension.
Additionally, because of differences due to mutation, each of the branches is also ``thick'' (high dimensional) when looked at closely.
Viewing this example as a low-dimensional subspace, as in PCA, is 
incorrect.

However, the local low-dimensionality can be exploited by looking on the right scales: a coarse scale in which the tree looks one-dimensional locally and a fine scale where the branch width matters.
We cover the tree with spheres (Box \ref{box:glos}) of radius $r_c$, where $r_c$ is on the order of the branch width; these spheres determine our clusters, and the number of them is the metric entropy of the tree \cite{tao2008product}.
Because all the points within a sphere are close to each other, they are highly 
redundant and can be encoded in terms of one another, saving space.

By the triangle inequality, in order to search for all points within distance $r$ of a query, we need only to look in nearby spheres with centers (i.e., representatives) within a distance $r+r_c$ of the query (Figure \ref{fig:dataStructure}D).
However, because each sphere has a radius comparable to branch width, the tree is locally one-dimensional on the coarse scale; that is, spheres largely tend to extend along the branches of the tree rather than in all directions.
We will call this property of local scaling the fractal dimension $d$ of the tree at the scale $r_c$ \cite{falconer1990fractal}, where $r_c$ is essentially our ruler size and $d=1$.
Thus, increasing the search radius for coarse search only linearly increases the number of points that need to be searched in a fine search.

A similar analysis holds in the more general case where $d \ne 1$.
The entropy-scaling frameworks we introduce can be expected to provide a boost to 
approximate search when fractal dimension $d$ of a dataset $D$ is low (i.e., close to 1) and metric
entropy $k$ is low.
Specifically, the ratio $\frac{|D|}{k}$ provides an estimate of the acceleration factor
for just the coarse search component compared to a full linear search of a database $D$.
Local fractal dimension around a data point can be computed by determining the
number of other data points within two radii $r_1$ and $r_2$ of that point;
given those point counts ($n_1$ and $n_2$, respectively), fractal dimension $d$
is simply $d=\frac{\log (n_2 / n_1)}{ \log (r_2 / r_1)}$.
Sampling this property over a dataset can provide a global average fractal 
dimension.
When we search a larger radius around a query, the number of points we encounter grows exponentially with the fractal dimension;
low fractal dimension implies that this growth will not obviate the gains provided by an entropy-scaling data structure.

More formally, given a database with fractal dimension $d$ and metric entropy $k$ at the scale $r_c$, we show in the Supplemental Methods that the time-complexity of similarity search on database $D$ for query $q$ with radius $r$ is
\begin{gather*}
    O\Bigg(
    \underbrace{k}_{\textrm{metric entropy}} +
    \overbrace{\left|B_D(q,r)\right|}^{\textrm{output size}}
    \underbrace{\left(\frac{r+2r_c}{r}\right)^d}_{\textrm{scaling factor}}
     \Bigg) .
\end{gather*}
Thus, for small fractal dimension and output size, similarity search is asymptotically linear in metric entropy.
Additionally, because the search has to look at only a small subset of the clusters, the clusters can be stored in compressed form, and only decompressed as needed, giving space savings that also scale with entropy.
The space-complexity scales with the sum of metric and information-theoretic entropy, rather than just metric entropy 
(Supplemental Methods: Theory).

\subsection{Practical application of entropy-scaling search}
We have presented the simplest such data to analyze for clarity of exposition.
However, real data is generally messier.
Sometimes the distance function is not a metric, so we lose the triangle inequality guarantee of 100\% sensitivity;
sometimes different distance functions can be used for the clustering versus search;
and sometimes even what counts as a distinct data point is not entirely clear without domain knowledge (for example, long genomic sequences might be better broken into shorter subsequences).

To show that entropy-scaling frameworks are robust to the variations presented by real data,
we explored a diversity of applications from three major biological ``big challenges of big data''---pharamaceuticals, metagenomics, and protein structure \cite{marx2013biology}.
We demonstrate that the general scheme results in order-of-magnitude improvements in running time in these different contexts, promising to enable new workflows for practitioners (e.g., fast first-pass computational drug screens and local analyses of sequencing data in remote field sites for real-time epidemic monitoring).
These applications are enabled by augmenting the framework with domain-specific distance functions in different stages of the process, as well as preprocessing to take advantage of domain-specific knowledge.
We expect that as long as the dataset exhibits both low entropy and low 
fractal dimension---and this is empirically true in biological systems---our 
entropy-scaling framework has the potential to achieve massive speedup 
over more na\"ive methods and significant speedup even over other highly 
optimized methods.

Source code for the applications discussed here is available at \url{http://gems.csail.mit.edu} and in the Supplemental Information.

\subsection{Application to high-throughput drug screening}

Chemogenomics is the study of drug and target discovery by using chemical
compounds to probe and characterize proteomic 
functions~\cite{bredel2004chemogenomics}.
Particularly in the field of drug discovery and drug repurposing, prediction 
of biologically active compounds is a critical task. 
Computational high-throughput screening can eliminate many compounds from 
wet-lab consideration, but even this screening can be time-consuming.
PubChem~\cite{bolton2008pubchem}, a widely-used repository of molecular compound 
structures, 
has grown greatly since 2008. 
In July 2007, PubChem contained 10.3 million compounds.
In October 2013, PubChem contained roughly 47 million compounds, while
in December 2014 it contained 61.3 million compounds.

We designed this compression and search framework around one of the standard 
techniques for high-throughput screening of potential drug compounds, the use 
of maximum common subgraph (MCS) to identify similar motifs among molecules \cite{cao2008maximum, rahman2009small}.
We introduce Ammolite, a method for clustering molecular databases such as 
PubChem, and for quickly searching for 
similar molecular structures in compressed space.
Ammolite demonstrates that entropy-scaling methods can be extended to data types that are not inherently sequence based.
Ammolite is a practical 
tool that provides approximately a factor of 150 speed-up with greater than 92\% accuracy compared to the popular small molecular subgraph detector (SMSD)~\cite{rahman2009small}.

An MCS-based search of molecule databases typically matches pairs of molecules by 
Tanimoto distance~\cite{rahman2009small}. 
Tanimoto distance obeys the triangle inequality and is more useful in the 
domain of molecular graphs than other
distance metrics such as graph distance \cite{bunke1998graph}.

To compress a molecule database, we project the space of small molecules onto a subspace by removing nodes and edges that do not participate in simple cycles
(Figure S2);
note that a molecule without cycles will collapse to a single node.
Clusters are exactly pre-images of this projection operator (i.e.,~all molecules that are isomorphic after simplification form a cluster).
Coarse search is performed by finding the MCS on this much smaller projection subspace.
This step increases speed by reducing both the required number of MCS operations 
and the time required for each MCS operation, which scales with the size of the molecule.
Further reduction in search time is accomplished by grouping clusters according
to size of the molecules within; because Tanimoto distance relies on molecule
size, clusters containing molecules significantly larger or smaller then the
query need not be searched at all.

The time required to cluster a large database such as PubChem is, nonetheless, significant; clustering the 306-GB PubChem required approximately 400 hours on a
12-core Xeon X5690 running at 3.47GHz, and required 128 GB RAM.
However, this database can easily be appended to as new molecules become available, and the clustering time can be amortized over future queries.
It is worth noting that the preprocessing of molecular graphs can cause the 
triangle inequality to be violated; while the distance function is a metric, the
clustering does not respect that metric.
Ammolite can be readily incorporated into existing analysis pipelines for 
high-throughput drug screening.

Our entropy-scaling framework can be applied to PubChem because it has both low fractal
dimension and low metric entropy.
In particular, we determined the mean local fractal dimension of PubChem to be 
approximately 0.2 in the neighborhood between 0.2 and 0.4 Tanimoto distance,
and approximately 1.9 in the neighborhood between 0.4 and 0.5.
The expected speedup is measured by the ratio of database size to metric entropy, which, for PubChem is approximately 11:1.
This is not taking into account the clustering according to molecule size, which
further reduces the search space.

Because SMSD is not computationally tractable on the entire PubChem database,
we benchmarked Ammolite against SMSD on a subset of 1 million molecules from PubChem.
Since SMSD's running time should scale linearly with the size of the database, we extrapolated the
running time of SMSD to the entire PubChem database.
Benchmarking Ammolite and SMSD required 60GB RAM and used 12 threads, although
Ammolite's search, used normally, requires $< 20$ GB RAM.
For these benchmarks, we used five randomly-chosen query molecules with at least two rings (PubChem IDs 1504670, 19170294, 28250541, 4559889, and 55484477), as well as five medically-interesting molecules chosen by hand (adenosine triphosphate [atp], clindamycin, erythromycin, teixobactin, and thalidomide).
We also used SMSD as a gold standard against which we measured Ammolite's recall.

Ammolite achieves an average of 92.5\% recall with respect to SMSD (Table~\ref{ammo1m}). This recall is brought down by one poorly-performing compound, PubChem ID 1504670, with only 62.5\% recall, but is otherwise over 80\%.
Furthermore, Ammolite's speed gains with respect to SMSD grow as the database grows (Table~\ref{ammo47m}).
\begin{table}
    \caption{Benchmarks of Ammolite vs. SMSD on databases of (a) 1 million molecules and (b) 47 million molecules (all of PubChem)}

\begin{subtable}{1\textwidth}
\caption{Ammolite benchmark on database of 1 million molecules}
\label{ammo1m}
\begin{tabular}{ccccc}
\hline
PubChem ID & SMSD (hours) & Ammolite (hours) & Speedup & Recall (\%) \\
\hline
5957 (atp) & 4.4 & 0.14 & 31 & 81\% \\
\hline
446598 (clindamycin) & 18.7 & 1.5 & 11.7 & 90\% \\
\hline
12560 (erythromycin) & 849.6 & 3.0 & 279.2 & 91\% \\
\hline
86341926 (teixobactin) & 618.5 & 2.3 & 265.5 & 100\% \\
\hline
5426 (thalidomide) & 48.9 & 0.81 & 60.4 & 100\% \\
\hline
1504670 & 8.1 & 0.8 & 10.3 & 62.5\% \\
\hline
19170294 & 31.3 & 0.8 & 39.7 & 100\% \\
\hline
28250541 & 43.3 & 4.8 & 9.0 & 100\% \\
\hline
4559889 & 108.8 & 2.7 & 41.0 & 100\% \\
\hline
55484477 & 23.3 & 2.5 & 9.1 & 100\% \\
\hline

\end{tabular}
\end{subtable}

\vspace{1em}

\begin{subtable}{1\textwidth}
\caption{Ammolite benchmark on entire PubChem database as of October 2013. See also Figure S2.}
\label{ammo47m}
\begin{tabular}{ccc}
\hline
PubChem ID & Ammolite (hours) & Speedup \\
\hline
5957 (atp) & 4.1 & 51.3 \\
\hline
446598 (clindamycin) & 28.4 & 14.5 \\
\hline
12560 (erythromycin) & 79.1 & 512.9 \\
\hline
86341926 (teixobactin) & 96.5 & 305.9 \\
\hline
5426 (thalidomide) & 29.2 & 80.0 \\
\hline
1504670 & 4.6 & 84.4 \\
\hline
19170294 & 6.0 & 247.4 \\
\hline
28250541 & 38.9 & 53.2\\
\hline
4559889 & 57.3 & 90.7 \\
\hline
55484477 & 35.5 & 31.4 \\
\hline
\end{tabular}
\end{subtable}
\end{table}

\subsection{Application to metagenomics}

Metagenomics is the study of genomic data sequenced directly from environmental
samples.
It has led to improved understanding of how ecosystems recover
from environmental damage~\cite{tyson2004community} and how the human gut responds 
to diet
and infection~\cite{david2014host}.
Metagenomics has even provided some surprising insights into disorders 
such as Autism Spectrum Disorder~\cite{macfabe2012short}.

BLASTX~\cite{altschul1990basic} is widely used in metagenomics to map
reads to protein databases such as KEGG~\cite{kanehisa2000kegg} and NCBI's 
NR~\cite{sayers2011database}.
This mapping is additionally used as a primitive in pipelines such as MetaPhlAn~\cite{segata2012metagenomic}, 
PICRUSt~\cite{langille2013predictive}, and MEGAN~\cite{huson2011integrative} to
determine the microbial composition of a sequenced sample.
Unfortunately, BLASTX's runtime requirements scale linearly with the product 
of the size of the full read dataset and the targeted protein database, and 
thus each year require exponentially more runtime to process the exponentially 
growing read data. 
These computational challenges are at present a barrier to widespread use of 
metagenomic data throughout biotechnology, which constrains genomic medicine 
and environmental genomics~\cite{frank2008gastrointestinal}.
For example, \citet{mackelprang2011metagenomic} reported that using BLASTX to map 246
million reads against KEGG required 800,000 CPU hours at a supercomputing 
center.

Although this is already a problem for major research centers, it is especially
limiting for on-site analyses in more remote locations.
In surveying the 2014 Ebola outbreak, scientists physically shipped samples on 
dry ice to Harvard for sequencing and analysis \cite{gire2014genomic}.
Even as sequencers become more mobile and can thus be brought on site, lack of fast Internet connections in remote
areas can make it impossible to centralize and expedite processing (viz., the cloud);
local processing on resource-constrained machines remains essential.
Thus, a better-scaling and accurate version of BLASTX raises the possibility of 
not only faster computing for large research centers, but also of performing
entirely on-site sequencing and desktop metagenomic analyses.

Recently, approaches such as RapSearch2~\cite{zhao2012rapsearch2} and 
Diamond~\cite{buchfink2014fast} have provided faster alternatives to BLASTX.
We have applied our entropy-scaling framework to the problem of 
metagenomic search and demonstrate MICA, a method whose software 
implementation provides an acceleration of DIAMOND by a factor of 3.5, and 
BLASTX by a factor of up to 3700.
This application illustrates the potential of entropy-scaling frameworks, while
providing a useful tool for metagenomic research.
It can be readily incorporated into existing analysis pipelines (e.g., for microbial 
composition analysis using MEGAN).
MICA clustering of the September 17, 2014 NCBI NR database (containing 49.3 million sequences) required 39 
hr on a 12-core Xeon X5690 running at 3.47GHz; it used approximately 84 GB of resident memory.

Our entropy-scaling framework can be applied to the NCBI's NR database because it, 
like PubChem, exhibits low fractal dimension and metric entropy.
We determined the mean local fractal dimension of the NCBI's NR database, using 
sequence identity of alignment as a distance function, to be approximately 1.6 
in the neighborhood between 70\% and 80\% protein sequence identity.
The ratio of database size to metric entropy, which gives an indicator of expected speedup,
is approximately 30:1.
Indeed, the notion that protein sequence space exhibits structure, 
and lends itself to clustering, has precedent~\cite{linial1997global}.

To evaluate the runtime performance of MICA, we tested it against
BLASTX, RapSearch2~\cite{zhao2012rapsearch2} and 
Diamond~\cite{buchfink2014fast}.
On five read sets (ERR335622, ERR335625, ERR335631, ERR335635, ERR335636) totalling 207,623 151-nucleotide (nt) reads from 
the American Gut Microbiome project, we found that MICA provides 
measurable runtime improvements over DIAMOND with no further loss in accuracy 
(Table~\ref{mgspeed}), and substantial runtime improvements over BLASTX.
Notably, the mean running time for BLASTX was 58,215 minutes, 
while MICA took an average of 15.6 minutes, a speedup of 3,724x.
MICA uses DIAMOND for its coarse search, and can use either DIAMOND or BLASTX
for its fine search.

We also evaluated MICA using BLASTX for both the coarse and the fine search; this approach performed slightly slower than DIAMOND, requiring an average of 89 min, though it was somewhat more accurate, at 95.9\% recall compared to DIAMOND's 90.4\% recall.
MICA using BLASTX for both coarse and fine searches relied on a query-side clustering (discussed in Supplemental Methods); we note that the time spent performing query-side clustering is included here; without query-side clustering, this variant of MICA takes 2,278 min, a speedup of 25x over BLASTX.

\begin{table}
\caption{(a) Running time and (b) accuracy of BLASTX, RapSearch2, DIAMOND, and MICA. Data set is the American gut microbiome project read sets ERR335622, ERR335625, ERR335631, ERR335635, ERR335636}
\begin{subtable}{1\textwidth}
\caption{Running time in minutes (standard deviation)}
\label{mgspeed}
\tabcolsep=0.08cm
\begin{tabular}{lllll}
\hline
   BLASTX & RapSearch2 & DIAMOND & MICA-DIAMOND & MICA-BLASTX \\
\hline
58215 (1561.8) & 206 (5.4) & 54 (1.1) & 15.6 (0.5) & 21.9 (1.7) \\
\hline
\end{tabular}
\end{subtable}

\begin{subtable}{1\textwidth}
\caption{Accuracy against BLASTX (standard deviation)}
\label{mgacc}
\tabcolsep=0.11cm
\begin{tabular}{lllll}
\hline
 RapSearch2 & DIAMOND & MICA-DIAMOND & MICA-BLASTX \\
\hline
79.5\% (1.63) & 90.4\% (3.10) & 90.4\% (3.10) & 90.4\% (3.10) \\
\hline
\end{tabular}
\end{subtable}
\end{table}

MICA accelerates DIAMOND with no further loss
in accuracy: 90.4\% compared to unaccelerated BLASTX Table~\ref{mgacc}.
Experiments validating accuracy treated BLASTX as a gold standard. 
Since MICA accelerates DIAMOND
using entropy-scaling techniques, false positives with respect to DIAMOND are 
not possible, but false negatives are.
We report as 
accuracy the fraction of BLASTX hits that are also returned by MICA.

DIAMOND's clever indexing and alphabet reduction provide excellent runtime performance already, though its running time still scales linearly with database size.
In contrast, as an entropy-scaling search, MICA will demonstrate greater
acceleration as database sizes grow~\cite{daniels2013compressive}.
Moreover, MICA can use standard BLASTX for its fine search, which allows the
user to pass arbitrary parameters to the underlying BLASTX call but which also comes at a small runtime penalty (40\% in our testing).
This option allows for additional BLAST arguments that DIAMOND does not support, such as XML output, which may be useful in some pipelines.
Thus, MICA with BLASTX may be suitable for a wider variety of existing analysis 
pipelines.

\subsection{Application to protein structure search}

The relationship between protein structure and function has been a subject of intense study for decades,
and this strong link has been used for the prediction of function from structure \cite{hegyi1999relationship}.
Specifically, given a protein of solved (or predicted) structure but unknown function, the efficient identification
of structurally similar proteins in the Protein Data Bank (PDB) is critical to function prediction.
Finding structural neighbors can also give insight into the evolutionary origins of proteins of interest \cite{yona1999protomap,nepomnyachiy2014global}.

One approach to finding structural neighbors is to attempt to align the query protein to all the entries in the PDB using a structural aligner, such as 
STRUCTAL \cite{subbiah1993structural}, ICE \cite{shindyalov1998protein}, or 
Matt \cite{menke2008matt}.
However, performing a full alignment against every entry in the PDB is prohibitively expensive, especially as the database grows.
To mitigate this, \citep{budowski2010fragbag} introduced the tool FragBag, which avoids performing full alignments but rather describes each protein as a
``bag of fragments,'' where each fragment is a small structural motif.
FragBag has been reported as comparable to structural aligners such as STRUCTAL or ICE,
and its bag-of-fragments approach
allows it to perform comparisons much faster than standard aligners.
Importantly for us, the bag of fragments is just a frequency vector, making
FragBag amenable to acceleration through entropy-scaling.

By first verifying that the local fractal dimension of PDB FragBag frequency vectors is low in most regimes ($d \approx 2-3)$, Figure S3), we are given reason to think that this problem is amenable to entropy-scaling search.
As an estimate of potential speedup, the ratio of PDB database size to metric 
entropy at for the chosen cluster radii is on average, $\sim$10:1.
We directly applied our entropy-scaling framework without any additional 
augmentation: esFragBag (entropy-scaling FragBag) is able to achieve an average
factor of 10 speedup of the highly-optimized FragBag with less than 0.2\% loss 
in sensitivity and no loss in specificity.

For this last example, we intentionally approach the application of entropy-scaling frameworks to FragBag in a blind manner,
without using any domain-specific knowledge.
Instead, we use the very same representation (bag of fragments) and distance functions (Euclidean and cosine distances)
as FragBag, coupled with a greedy k-centers algorithm to generate the clustered representation.
Note that this is in contrast to MICA and Ammolite, which both exploit domain knowledge to further improve performance.
Thus, esFragBag only involves extending an existing codebase with new database generation and similarity search functions.

\begin{figure}[p]
    \centering
    \vspace{-10em}
    \centerline{
    \begin{subfigure}[b]{3.0in}
        \caption{Cosine distance}
        \label{fig:fragbag_cosine}
        \includegraphics[width=1\textwidth]{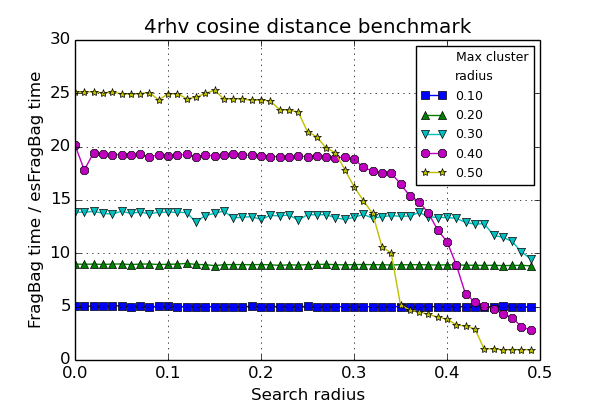}
    \end{subfigure}%
    \begin{subfigure}[b]{3.0in}
        \caption{Euclidean distance}
        \label{fig:fragbag_euclid}
        \includegraphics[width=1\textwidth]{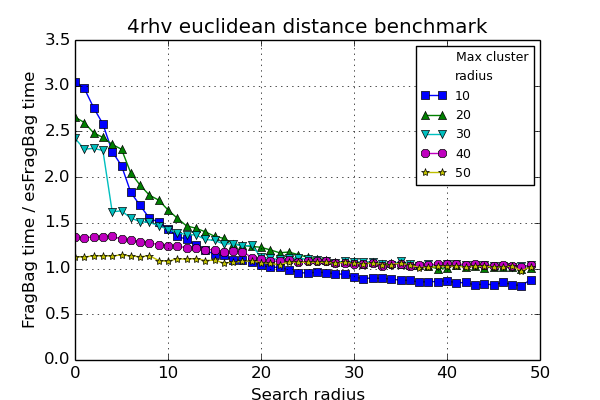}
    \end{subfigure}
    }
    \centerline{
    \begin{subfigure}[b]{3.0in}
        \includegraphics[width=1\textwidth]{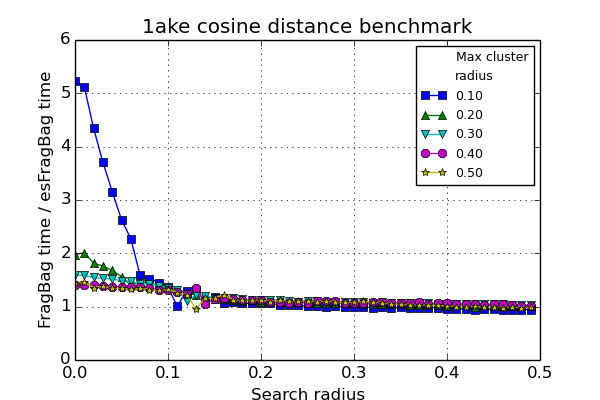}
    \end{subfigure}%
    \begin{subfigure}[b]{3.0in}
        \includegraphics[width=1\textwidth]{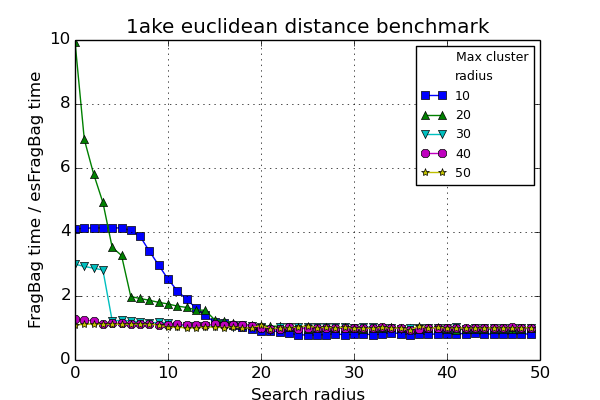}
    \end{subfigure}
    }
    \centerline{
    \begin{subfigure}[b]{3.0in}
        \includegraphics[width=1\textwidth]{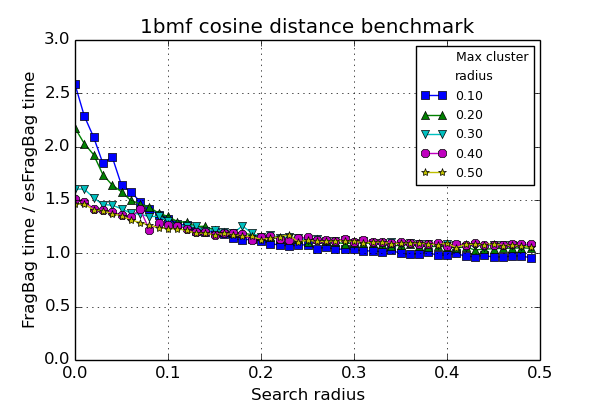}
    \end{subfigure}%
    \begin{subfigure}[b]{3.0in}
        \includegraphics[width=1\textwidth]{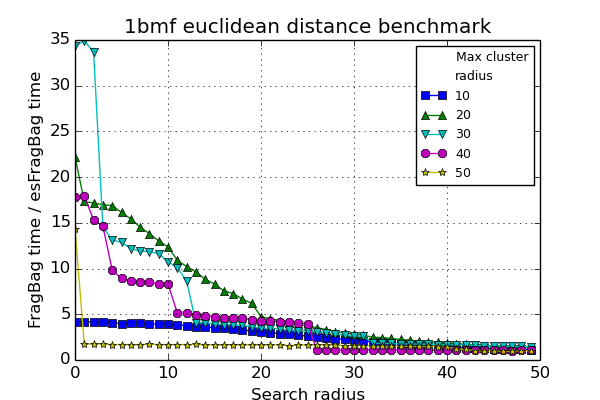}
    \end{subfigure}
    }
    \centerline{
    \begin{subfigure}[b]{3.0in}
        \includegraphics[width=1\textwidth]{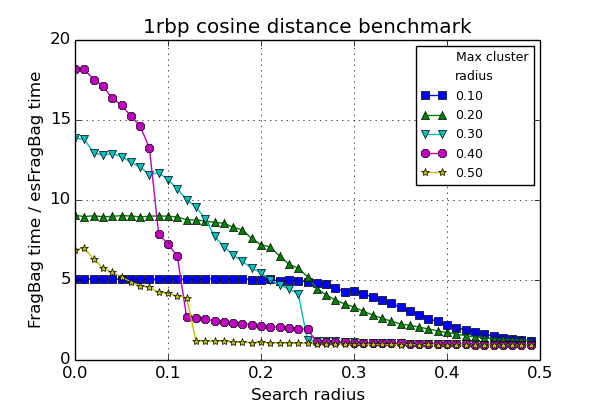}
    \end{subfigure}%
    \begin{subfigure}[b]{3.0in}
        \includegraphics[width=1\textwidth]{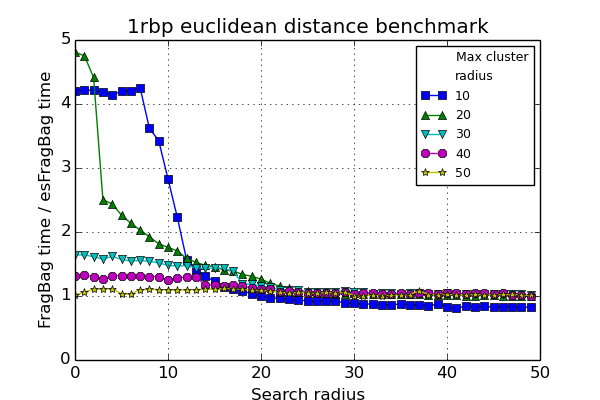}
    \end{subfigure}
    }
    \caption{Scaling behavior of esFragBag. EsFragBag benchmarking data with parameters varied until the acceleration advantage of esFragBag disappears. As search radius increases, the fraction of the database returned by the coarse search increases, ultimately returning the whole database. Unsurprisingly, when returning the whole database in the coarse search results, there are no benefits to using entropy-scaling frameworks. (a) Cosine distance gives on the whole better acceleration, but results in $>99.8\%$ sensitivity, whereas (b) Euclidean distance as a metric is guaranteed by the Triangle Inequality to get $100\%$ sensitivity.}
    \label{fig:fragbag}
\end{figure}

We investigate the increases in speed resulting from directly applying the entropy-scaling framework for both Euclidean and cosine distances and found the acceleration is highly dependent on both the search radius and cluster radius (Figure \ref{fig:fragbag}).
For cosine distance, we generated databases with maximum cluster radii of 0.1, 0.2, 0.3, 0.4, and 0.5.
Then, for each query protein from the set \{\texttt{4rhv}, \texttt{1ake}, \texttt{1bmf}, \texttt{1rbp}\} (identified by PDB IDs), we ran both na\"ive and accelerated similarity searches with radii of $0.02i, \forall i \in \{0,\ldots,49\}$.
This test was repeated 5 times for each measurement, and the ratio of average accelerated vs na\"ive times is shown in Figure \ref{fig:fragbag_cosine}.

For Euclidean distance, we generated databases with maximum cluster radii of 10, 20, 25, 50, and 100.
Again, for each query protein drawn from the same set, we compared the average over five runs of the ratio of average accelerated versus na\"ive times (Figure \ref{fig:fragbag_euclid}).
The cluster generation required anywhere from 65 to 23,714 seconds, depending on the choice of radii (See table \ref{tab:esfragbag_clustering}) and no more than a small constant ($<3$) times as much memory as it takes to simply load the PDB database (no more than 2 GB RAM).
Clustering used 20 threads on a 12-core Xeon X5690, while search used only one thread.
\begin{table}
\caption{Cluster generation time for esFragBag}
\label{tab:esfragbag_clustering}
\begin{subtable}{1\textwidth}
    \centering
\caption{Cosine distance:}
\begin{tabular}{c|ccccc}
\hline
\textbf{radius} & 0.1 & 0.2 & 0.3 & 0.4 &  0.5 \\
\hline
\textbf{time (s)} & 21,037 & 11,088 & 7,409 & 5,288 & 3,921 \\
\hline
\end{tabular}
\end{subtable}

\vspace{1em}
\begin{subtable}{1\textwidth}
    \centering
\caption{Euclidean distance:}
\begin{tabular}{c|ccccc}
\hline
\textbf{radius} & 10 & 20 & 30 & 40 & 50 \\
\hline
\textbf{time (s)} & 23,714 & 3,062 & 483 & 144 & 65 \\
\hline
\end{tabular}
\end{subtable}
\end{table}

Not only is the acceleration highly dependent on both the search radius $r$ and the maximum cluster radius $r_c$,
but the choice of query protein also affects the results.
We suspect that this effect is due to the geometry of protein fragment frequency space being very ``spiky'' and ``star-like''.
Proteins that are near the core (and thus similar to many other proteins) show very little acceleration when our framework is used because the majority of the database is nearby, whereas proteins in the periphery have fewer neighbors and are thus found much more quickly.
Changing the maximum cluster radius effectively makes more proteins peripheral proteins, but at the cost of overall acceleration.

Naturally, as the search radius expands, it quickly becomes necessary to compare against nearly the entire database, destroying any acceleration.
For the cosine space in particular, note that the maximum distance between any two points is $1$, so once the coarse search radius of $r+r_c \ge 1.0$, there cannot ever be any acceleration as the fine search encompasses the entire database.
Similarly, once the coarse search encompasses all (or nearly all) the clusters in Euclidean space, the acceleration diminishes to a factor 1, and the overhead costs make the entropy-scaling framework perform worse than a na\"ive search.
However, as we are most interested in proteins that are very similar to the query, the low-radius behavior is of primary interest.
In the low-radius regime, esFragBag demonstrates varying though substantial acceleration (2-30x, averaging $>$10x for both distance functions for the proteins chosen) over FragBag.

It is instructive to note that because of the very different geometries of Euclidean vs cosine space, acceleration varies tremendously for some proteins, such as \texttt{4rhv} and \texttt{1bmf}, which display nearly opposite behaviors.
Whereas there is nearly 30x acceleration for \texttt{4rhv} in cosine space for low radius, and the same for \texttt{1bmf} in Euclidean space, neither achieves better than $\sim$ 2.5x acceleration in the other space.

Finally, while Euclidean distance is a metric---for which the triangle inequality guarantees 100\% sensitivity---cosine distance is not.
Empirically, however, for all of the queries we performed, we achieve $> 99.8\%$ sensitivity (Table \ref{tab:fragbag_cosine_sensitivity}).

\begin{table}
    \centering
    \caption{Average sensitivity of esFragBag compared to FragBag when using cosine distance for the trials described in Figure \ref{fig:fragbag_cosine}. This table averages the sensitivities for each choice of search radii $\{0, 0.01, \ldots, 0.49\}$. (NB: no analogous table is given for Euclidean distance as the Triangle Inequality ensures perfect recall).}
    \label{tab:fragbag_cosine_sensitivity}
    \begin{tabular}{|c|cccc|}
        \hline
        \backslashbox{Cluster radii}{Query protein}  & 4rhv & 1ake & 1bmf & 1rbp \\
        \hline
        0.10  & 1  & 0.999840     & 0.998490 & 0.999950  \\
        0.20  & 1  & 0.999918     & 0.999001 & 0.999978  \\
        0.30  & 1  & 0.999926     & 0.999649 & 1  \\
        0.40  & 1  & 0.999974     & 0.999796 & 1  \\
        0.50  & 1  & 0.999984     & 0.999934 & 1  \\
        \hline
    \end{tabular}
\end{table}

\section{Application to other domains}

We anticipate that our entropy-scaling approach will be useful to other kinds of
biological data sets; applying it to new data sets will require several steps.
Here we provide a ``cookbook'' for applying our entropy-scaling framework to
a new data set.
Given a new data set, we first define what the high-dimensional space is.
For metagenomic sequence data, it is the set of enumerable protein sequences up
to some maximum length, while for small-molecule data, it is the set of
connected chemical graphs up to some maximum size, and for protein structure
data (using the FragBag model) it is the set of ``bag-of-words'' frequency 
vectors of length 400.

Given the high dimensional space, we determine how database entries map
onto points (for example, in the case of MICA, they are greedily broken into subsequences with a minimum length).
Next, clustering can be implemented; a simple greedy clustering may suffice 
(as for esFragBag) but clustering of sequence data may be dramatically 
accelerated by using BLAST-style seed-and-extend matching (as used in MICA).
Finally, coarse and fine search can be implemented; in many cases, existing
tools may be used ``out of the box,'' as with esFragBag and MICA.
With MICA, we note that coarse search by default uses
DIAMOND, while fine search provides a choice of DIAMOND or BLASTX.
With Ammolite, we used the SMSD library, but incorporated it into our own search
tool.

\section{Discussion}

We have introduced an entropy-scaling framework for accelerating approximate search, allowing search on large omics datasets to scale, even as those datasets grow exponentially.
The primary advance of this framework is that it bounds both time and space as functions of the dataset entropy (albeit using two different notions of entropy: metric entropy bounds time, while information-theoretic entropy bounds space).
We proved that runtime scales linearly with the entropy of the database, but we also show (Supplemental Methods: Theory) that under certain additional constraints, this entropy-scaling framework permits a compressed representation on disk.
This compression is particularly applicable in the case of metagenomic analysis, where the collection of read data presents a major problem for storage and transfer.
Although we did not optimize for on-disk compression in any of our applications, choosing instead to focus on search speed, implementing this compression is feasible using existing software tools and libraries such as Blocked GZip (BZGF); each cluster would be compressed separately on disk.

Furthermore, we have justified and demonstrated the effectiveness of this 
framework in
three distinct areas of computational molecular biology, providing the
following open-source software: Ammolite for
small-molecule structure search, MICA for metagenomic analysis, and esFragBag for protein structure search.
All of our software is available under the GNU General Public License, and not only can the tools we are 
releasing be readily plugged into existing pipelines, but the code and 
underlying methods can also be easily incorporated into the original 
software that we are accelerating.

The reason for the speedup is the combination of low fractal dimension and low metric entropy.
Low fractal dimension ensures that runtime is dominated by metric entropy.
The size of the coarse database provides an estimate of metric entropy.
Furthermore, we can directly measure the local fractal dimension of the database by sampling points from the database and looking at the scaling behavior of the number of points contained in spheres of increasing radii centered on those sampled points.
We have shown that for three domains within biological data science, metric entropy, and fractal dimension are both low.

As discussed in the theoretical results, although the data live locally on a 
low dimension subspace, the data are truly high-dimensional globally.
At small scales, biological data often lives on a low-dimensional polytope~\cite{hart2015inferring}.
However, omics data are, by nature, comprehensive and include not just one but many such polytopes.
Although each polytope can be individually projected onto a subspace using techniques such as PCA, the same projection cannot be used for all the polytopes at once because they live on different low-dimensional subspaces.
Furthermore, as is the case with genomes, the low-dimensional polytopes are also often connected (e.g., through evolutionary history).
Thus, collections of local projections become unwieldy.
By using our clustering approach, we are able to take advantage of the existence of these low-dimensional polytopes for accelerated search without having to explicitly characterize each one.

A hierarchical clustering approach, rather than our flat clustering, has the
potential to produce further gains~\cite{loh2012compressive}.
We have taken the first steps in exploring this idea here; the
molecule size clustering in Ammolite can be thought of as an initial version of a multi-level or hierarchical clustering.

Entropy-scaling search is related to succinct, compressed, and opportunistic data structures, such as the compressed suffix array, the FM-index, and the sarray \cite{grossi2005compressed,ferragina2000opportunistic,conway2011succinct}.
However, these solve the problem of theoretically fast and scalable pattern matching (Box \ref{box:glos}), whereas we solve, theoretically and practically, the much more general similarity search problem.
An entropy-scaling search tree is also related to a metric ball tree \cite{uhlmann1991satisfying}, although with different time complexity.
Querying a metric ball tree requires $O(\log n)$ time, assuming the relatively uniform distribution of data points in a metric space.
This distribution differs from the non-uniform distribution under which entropy-scaling search behaves well.
In future work, we will investigate further acceleration of coarse search by applying a metric ball tree to the cluster representatives themselves; this approach may reduce the coarse search time to $O(\log k)$.
This step, too, can be thought of as an additional level of clustering.

Other metric search trees can also be found in the database literature \cite{zezula2006similarity}, although, to our knowledge, they have not been explicitly applied to biological data science.
The closest analogue to entropy-scaling search trees is the M-tree \cite{ciaccia1997deis,ciaccia1998cost}, which resembles a multi-level variation of our entropy-scaling search trees.
However, the M-tree time-complexity analysis \cite{ciaccia1998cost} does not have a nice closed form and is more explicitly dependent on the exact distribution of points in the database.
By using and combining the concepts of metric entropy and fractal dimension for our analysis, we are able to give an easier to understand and more intuitive, if somewhat looser, bound on entropy-scaling search tree complexity.

Entropy-scaling frameworks have the advantage of becoming proportionately faster and space efficient with the size of the available data.
Although the component pieces (e.g., the clustering method chosen) of the framework can be either standard (as in esFragBag) or novel (as in Ammolite), the key point is that these pieces are used in a larger framework to exploit the underlying complex structure of biological systems, enabling massive acceleration by scaling with entropy.
We have demonstrated this scaling behavior for common problems drawn from metagenomics, cheminformatics, and protein structure search, but the general strategy can be applied directly or with simple domain knowledge to a vast array of other problems faced in data science.
We anticipate that entropy-scaling frameworks should be applicable beyond the life sciences, wherever physical or empirical laws have constrained data to a subspace of low entropy and fractal dimension.

\section{Methods}

\subsection{Ammolite small molecule search}
Ammolite's clustering approach relies on structural similarity.
We augmented the entropy-scaling data structure by using a clustering scheme based on molecular structural motifs instead of a distance function.
Each molecule is ``simplified'' by removing nodes and edges that do not
participate in simple cycles.
Clusters are formed of molecules that are isomorphic after this simplification
step.
Each cluster can then be represented by a single molecular structure, along 
with pointers to ``difference sets''  between that structure and each of the 
full molecules in the cluster it represents.
For both coarse and fine search, we use the Tanimoto distance metric, defined as
\[d(G_1,G_2) = 1 - \frac{ |mcs(G_1,G_2)| }{|G_1|+|G_2|-|mcs(G_1,G_2)|},\]
where $mcs$ refers to the maximum common subgraph of two chemical graphs. 

The coarse search is performed in compressed space, by searching 
the coarse database with the goal of identifying possible hits.
The query molecule is simplified in exactly the same manner as 
the molecular database during clustering, and this transformed query graph is 
matched against the coarse database.
To preserve sensitivity, this coarse search is performed with a permissive 
similarity score.
Any possible hits---molecular graphs from the coarse database whose MCS to 
the transformed query molecule was within the similarity score threshold---are 
then reconstructed by following
pointers to the removed atom and bond information and recreating the 
original molecules.
Since the Tanimoto distance is used, we can bound the size of candidate 
molecules based on the size of the query molecule and the desired Tanimoto 
cutoff.
Thus, a second level of clustering, at query time, based on molecule size, 
allows further gains in runtime performance.
Finally, the fine search is performed against these decompressed possible 
hits that are within the appropriate size range based on the Tanimoto distance
cutoff.

\subsection{MICA metagenomic search}
CaBLASTX's clustering approach relies on sequence similarity.
We augmented the entropy-scaling data structure by using
different distance functions for clustering and search.
For clustering, we rely on sequence identity, while for search, we use the
E-value measure that is standard for BLAST.
All benchmarks were performed with an E-value of $10^{-7}$. For coarse search, MICA uses the DIAMOND argument \texttt{--top 60} in order to return all queries
with a score within 60\% of the top hit.
When MICA was tested using BLASTX for coarse search, it used an E-value of 
1000.
This seemingly surprisingly large coarse E-value is used because E-values are poorly behaved for short sequences; in sensitivity analysis, coarse E-values of 1 and 10 exhibited recall below 10\%, and an E-value of 100 exhibited recall below 60\%.
Furthermore, during clustering (compression), we apply a preprocessing step that
identifies subsequences to be treated as distinct points in the database.
We apply a reversible alphabet reduction to the
protein sequences, which projects them into a subspace (Supplemental Methods).

When applied to high-coverage next-generation sequencing queries, caBLASTX can also perform clustering on the reads (Supplemental Methods).
In this instance, coarse search is performed by matching each representative query with a set of representative database entries.
Fine search then matches the original queries within each cluster with the candidate database entries resulting from the coarse search.

\subsection{esFragBag protein structure search}
In FragBag, the bag of fragments is essentially
a term frequency vector representing the number of occurrences of each structural motif within the protein.
FragBag turns out to be amenable to acceleration using an entropy-scaling data structure because much of the computation is spent in doing a similarity search on that frequency vector.

For the cluster generation, we trivially used a na\"ive randomized greedy 2-pass approach.
First, all proteins in the Protein Data Bank were randomly ordered.
Then in the first pass, proteins were selected as cluster centers if and only if they were not within a user-specified Euclidean distance $r_c$ from an existing center (i.e., the first protein is always selected, and the second if further away than $r_c$ from the first, etc.).
Recall that this generation of cluster centers is the same as the one used to generate covering spheres in Figure \ref{fig:tree};
the covering spheres were overlapping, but we assign every protein uniquely to a single cluster by assigning to the nearest cluster center in the second pass.

Similarity search here was performed exactly as described in the section ``Entropy-Scaling Similarity Search'', with no modification.
For a given search query $q$ and search radius $r$,
a coarse search was used to find all cluster centers within distance $r+r_c$ of $q$.
Then, all corresponding clusters were unioned into a set $F$.
Finally, a fine search was performed over the set $F$ to find all proteins within distance $r$ of $q$.

\section{Author Contributions}
Y.W.Y., N.M.D., and B.B. conceived the project.
Y.W.Y., N.M.D., and B.B. developed the theoretical analyses.
N.M.D. and D.C.D. implemented and benchmarked MICA.
D.C.D. implemented and benchmarked Ammolite, with help from N.M.D. and Y.W.Y.
Y.W.Y. implemented and benchmarked esFragBag, with help from N.M.D.
B.B. guided all research and provided critical advice on the study.
Y.W.Y., N.M.D. and B.B. wrote the manuscript.

\section{Acknowledgments}
Y.W.Y. is supported by a Hertz Foundation fellowship.
N.M.D. and B.B. are supported by NIH GM108348.
We thank Andrew Gallant for his implementation of Fragbag.
We thank Joseph V. Barrile for graphic design.
We thank Jian Peng for suggesting high-throughput drug screening as an application.

\bibliographystyle{model5-names-noitalic}
\bibliography{main}

\end{document}


\renewcommand\thefigure{S\arabic{figure}}
\setcounter{figure}{0}
\renewcommand\thetable{S\arabic{table}}
\setcounter{table}{0}


\begin{figure}[hbp]
    \centering
    \centerline{
    \begin{subfigure}[b]{4in}
        \includegraphics[width=1\textwidth]{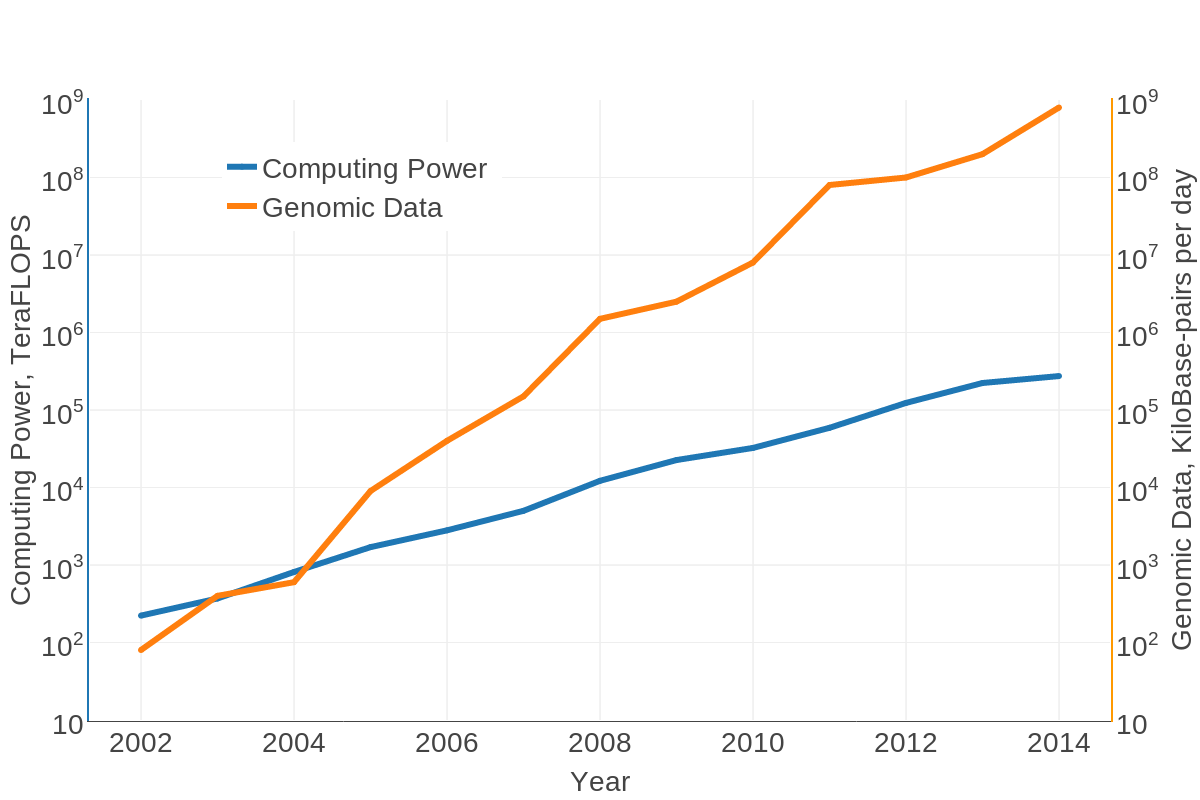}
        \caption{}
    \end{subfigure}%
    \begin{subfigure}[b]{4in}
        \includegraphics[width=1\textwidth]{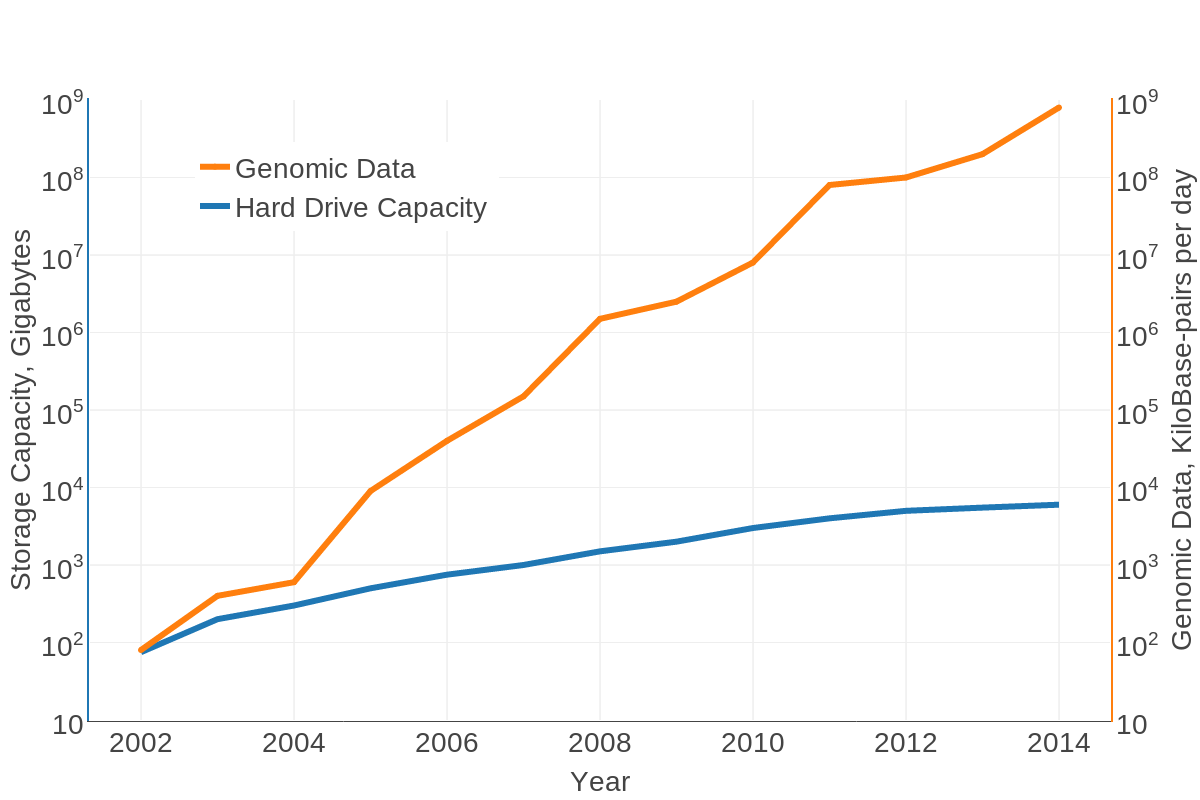}
        \caption{}
    \end{subfigure}}
    \caption{Genomic data available has grown at a faster exponential rate than computer processing power and disk storage.
    These plots represent, on a log scale, the daily growth in sequence data 
    from GenBank along with (a) the combined computing power (in TeraFLOPs) of 
    the Top 500 Supercomputer list, and (b) the largest commercially-available 
    hard disk drives.}
    \label{fig:expdata}
\end{figure}

\begin{figure}[hbp]
    \centering
    \includegraphics[width=1\textwidth]{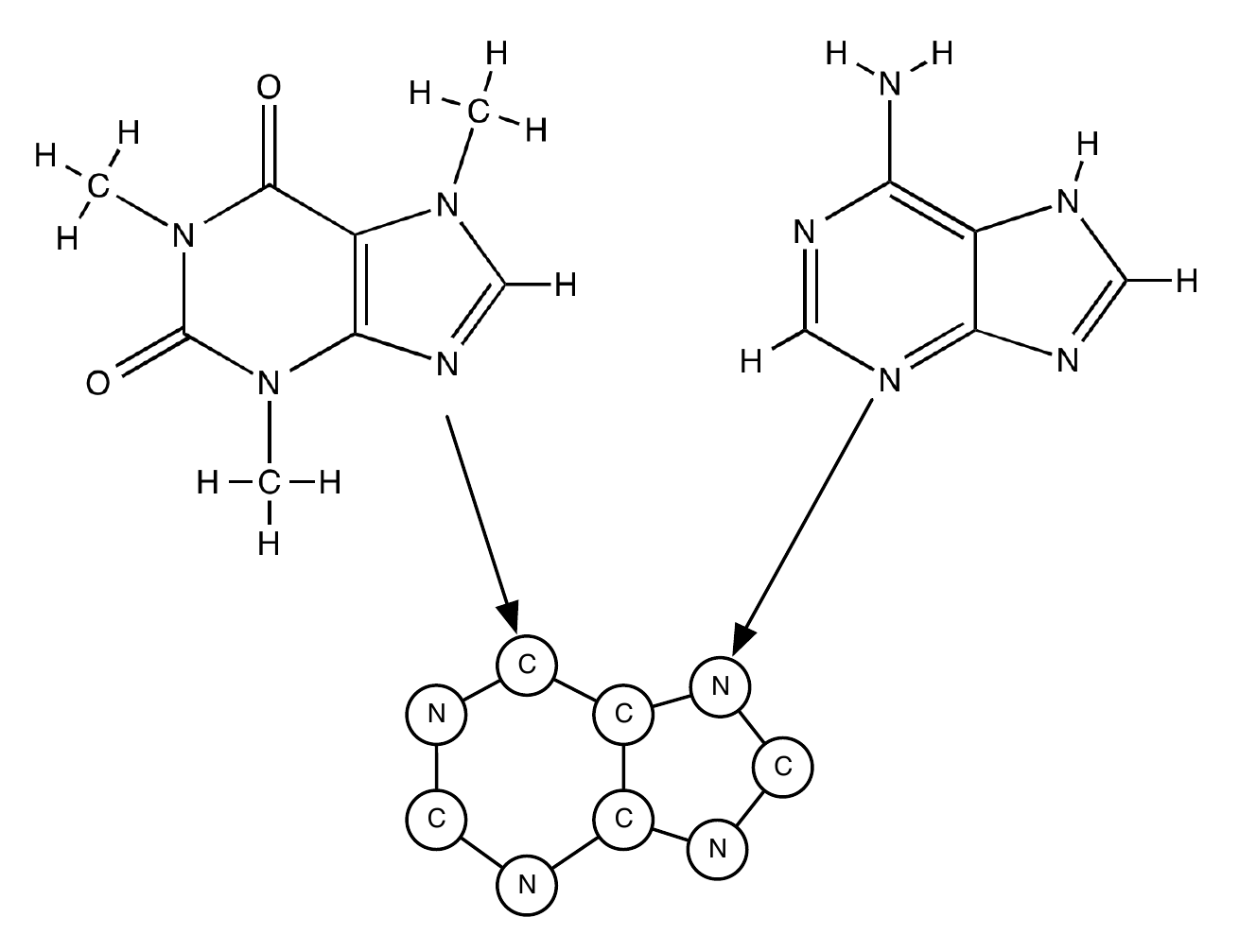}
    \caption{\textbf{(Related to Table 1b)} Ammolite's preprocessing during the clustering phase. Ammolite 
    removes nodes and edges that do not participate in simple cycles, and 
    treats all edges as simple, unlabeled edges. In this example, both caffeine 
    and adenine become a purine-like graph structure. Note that the resulting 
    graph has no implicit hydrogens.
    }
    \label{fig:ammolite}
\end{figure}

\begin{figure}[p]
    \centering
    \vspace{-10em}
    \centerline{
    \begin{subfigure}[b]{3.2in}
        \caption{Cosine distance}
        \label{fig:fragbag_fractal_cosine}
        \includegraphics[width=1\textwidth]{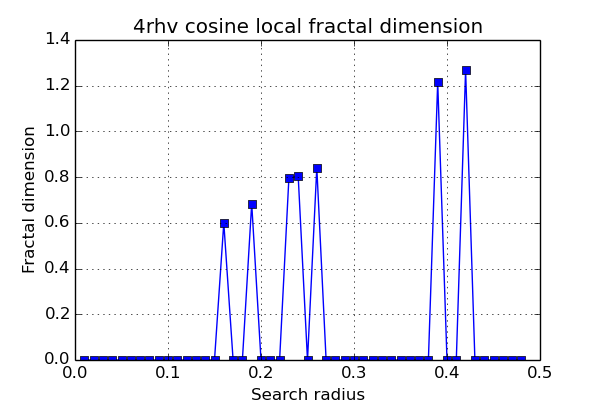}
    \end{subfigure}%
    \begin{subfigure}[b]{3.2in}
        \caption{Euclidean distance}
        \label{fig:fragbag_fractal_euclid}
        \includegraphics[width=1\textwidth]{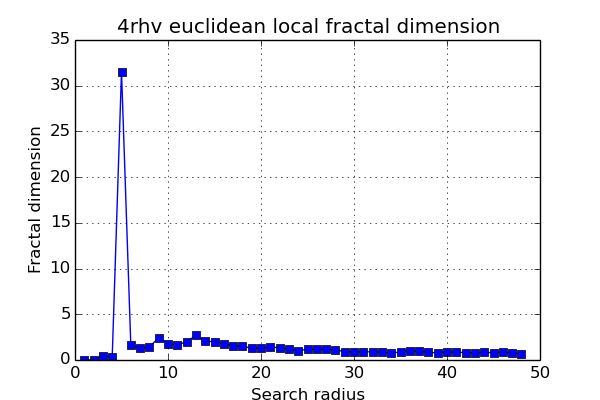}
    \end{subfigure}
    }
    \centerline{
    \begin{subfigure}[b]{3.2in}
        \includegraphics[width=1\textwidth]{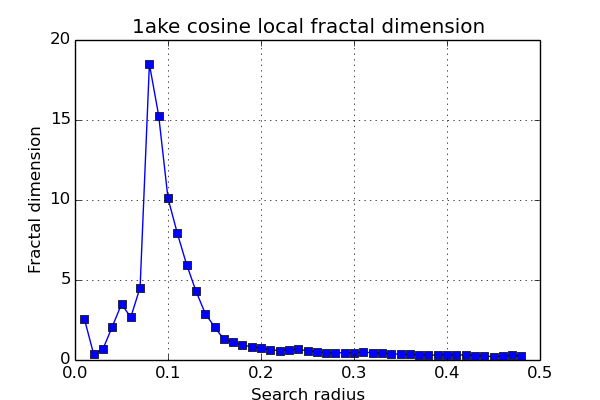}
    \end{subfigure}%
    \begin{subfigure}[b]{3.2in}
        \includegraphics[width=1\textwidth]{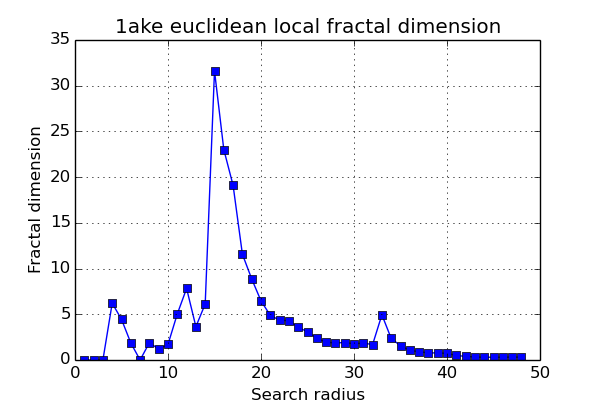}
    \end{subfigure}
    }
    \centerline{
    \begin{subfigure}[b]{3.2in}
        \includegraphics[width=1\textwidth]{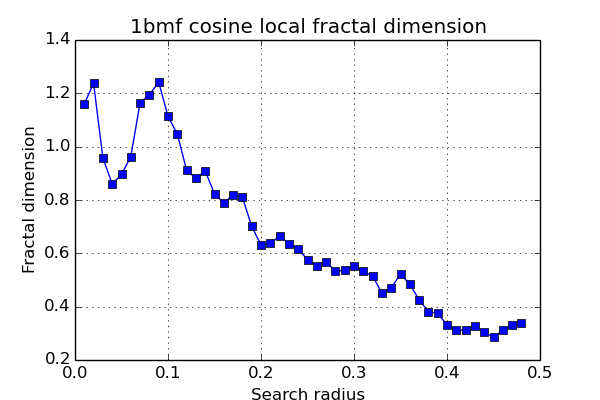}
    \end{subfigure}%
    \begin{subfigure}[b]{3.2in}
        \includegraphics[width=1\textwidth]{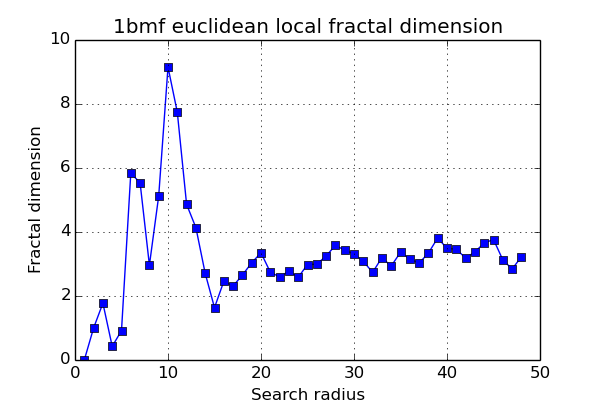}
    \end{subfigure}
    }
    \centerline{
    \begin{subfigure}[b]{3.2in}
        \includegraphics[width=1\textwidth]{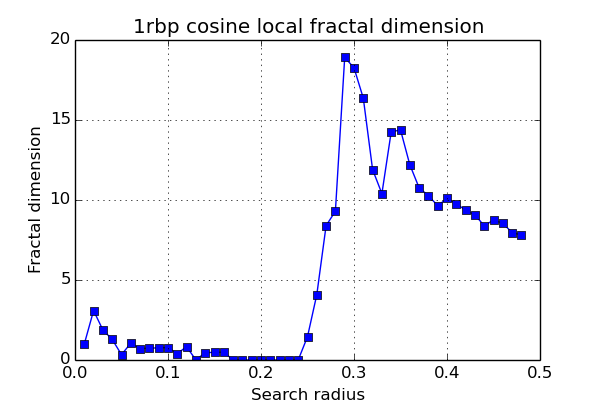}
    \end{subfigure}%
    \begin{subfigure}[b]{3.2in}
        \includegraphics[width=1\textwidth]{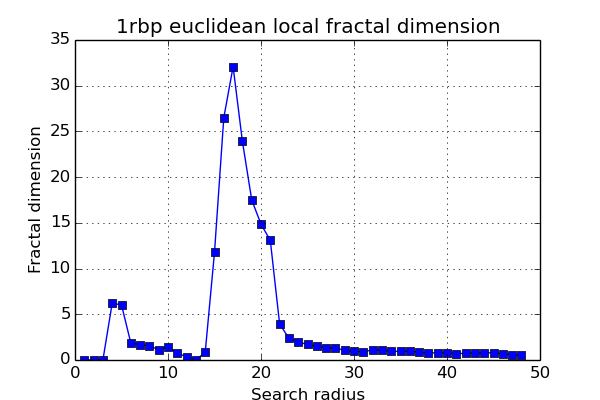}
    \end{subfigure}
    }
    \caption{\textbf{(Related to Figure 3)} Local fractal dimension at different scales for the space of PDB FragBag frequency vectors. Each data point is defined by dimension $d = \frac{\log(n_2/n_1)}{\log(r_2/r_1)}$, where $n_1, n_2$ are the number of similarity search hits within radius respectively $r_1, r_2$, and $r_2 - r_1$ is the increment size of 0.01 for cosine distance and 1 for euclidean distance. In most regimes, local fractal dimension is consistently low, except for a large spike when radius expands to include the central cluster of proteins. esFragBag achieves the most acceleration when both output size is small and we remain in a low fractal dimension regime.}
    \label{fig:fragbag_fractal}
\end{figure}

\section{Supplemental Methods}

\subsection{Theory}
\subsubsection{Time-complexity}
We introduced the definition of the entropy-scaling similarity search data structure in Figure 1.
For ease of analysis, we will work in a high-dimensional metric space and consider the database as a set $D$ of $n$ unique points in that metric space.
Define $B_S(q,r) = \{ p \in S : ||q-p||<r \}$. The similarity search problem is thus to compute $B_D(q,r)$ for a query $q$ and radius $r$.
Note however that the metricity requirement is needed only for a 100\% sensitivity guarantee; other distance functions can be used, but result in some loss in sensitivity.
However, regardless of the distance function chosen, there cannot be a loss of
specificity; false positives will never be introduced because the fine search is just the original search function on a smaller subset of the database.

A set $C$ of $k$ cluster centers are chosen such that no cluster has radius greater than a user-specified parameter $r_c$ and no two cluster centers are within distance $r_c$ of one another.
The data structure then clusters the points in the set by assigning them to their nearest cluster center.
Overloading notation a bit, we will identify each cluster with its center, so $C$ is also the set of clusters.
For a given similarity search query for all items within distance $r$ of a query $q$, this data structure breaks the query into coarse and fine search stages.
The coarse search is over the list of cluster centers, returning $B_C(q,r + r_c)$.
Let \[\displaystyle F = \bigcup_{c \in B_C(q,r+r_c)} c , \] the union of all the returned clusters.
By the Triangle Inequality, $B_D(q,r) \subseteq F$, which combined with $F \subseteq D$ implies that $B_F(q,r) = B_D(q,r)$.
Thus, a fine search over the set $F$ will return all items within radius $r$ of $q$.

Note that we require the metricity requirement only for the Triangle Inequality.
It turns out that many interesting distance functions are not metrics, but still almost satisfy the Triangle Inequality, which is nearly sufficient.
More precisely, if a fraction $\alpha$ of the triples in $S$ do not satisfy the Triangle Inequality, then in expectation, we will have sensitivity $1 - \alpha$.
As shown in the results, empirically, this loss in sensitivity appears to be low and can likely be ameliorated by increasing the coarse search radius.

Provided the fractal dimension of the database is low, and given a few technical assumptions on the distribution of points, this data structure allows for similarity search queries in time roughly linear in the metric entropy of the database.
Additionally, without increasing the asymptotic time-complexity, this data structure can also be stored in an \textit{information theoretic} entropy-compressed form.

Note that entropy-scaling data structures are distinct from both succinct data structures and compressed data structures.
Succinct data structures are ones that use space close to the information-theoretic limit in the worst case while permitting efficient queries; i.e.
succinct data structures do not depend on the actual entropy of the underlying data set, but have size-dependence on the potential worst-case entropy of the data set \cite{jacobson1988succinct}.
Compressed (and opportunistic) data structures, on the other hand, bound the amount of the space used by the entropy of the data set while permitting efficient queries \cite{grossi2005compressed, ferragina2000opportunistic}.
Entropy-scaling data structures are compressed data structures, but are distinct, as
unlike entropy-scaling data structures, compressed data structures do not measure time-complexity in terms of metric entropy.
Additionally, existing compressed data structures such as the compressed suffix array and the FM-index are designed for the problem of pattern matching \cite{grossi2005compressed, ferragina2000opportunistic}.
While related to similarity search, pattern matching does not admit as general of a notion of distance as the similarity search problem.
While compressed sensing has also been applied to the problem of finding a representative set of genes for a collection of expression samples~\cite{prat2011recovering}, compressed sensing is distinct from entropy-scaling data structures.

\textbf{The primary advance of entropy-scaling data structures is that they bound both space and time as functions of the data set entropy (albeit using two different notions of entropy).}

\subsubsection{Unifying different notions of fractal dimension}
In this paper, we make use of two different notions of fractal dimension, including the more intuitive formulation in the main paper and a more classical definition in this following section.
In order to both unify these different notions of fractal dimension and to allow us to prove our complexity bounds, we make a technical assumption about the self-similarity of the data set.
Roughly speaking, we require that the density of points be similar throughout the data set.

More precisely, for a given radius, consider the number of neighbors about any point for that radius, which we will call the density around a point.
We assume that for any given radius, the densities around all points are bounded within a constant multiplicative factor $\gamma$ of one another.
This particular technical assumption is likely stronger than we need, but makes the following arguments much more convenient;
we conjecture but do not prove here that so long as the distribution of densities has low variance, all our statements below also hold with high probability.
We give special thanks to one of our commenters for having brought this subtlety to our attention.

\subsubsection{Complexity bounds}
We first define the concept of metric entropy and entropy dimension in the classical manner:
\begin{definition}[\cite{tao2008product} Definition 6.1] 
    Let $X$ be a metric space, let $D$ be a subset of $X$, and let $\rho>0$ be a radius.
    \begin{itemize}
        \item The \textit{metric entropy} $N_\rho(D)$ is the fewest number of points $x_1, \ldots, x_n \in D$ such that the balls $B(x_1,\rho), \ldots B(x_n,\rho)$ cover $D$.
    \end{itemize}
\end{definition}
\begin{definition}[\cite{falconer1990fractal}]
    The Hausdorff dimension of a set $D$ is given by 
\[
    \dim_{\text{Hausdorff}}(D) := \lim_{\rho \to 0} \frac{\log N_\rho(D)}{\log 1/\rho}
\]
\end{definition}
Unfortunately, as $D$ is a finite, discrete, set, the given definision always gives $\dim_{\text{Hausdorff}}(D) = 0$.
However, we are only interested in scaling behaviors around large radii, so instead we use:
\begin{definition}
    Define fractal dimension $d$ of a set $D$ at a scale $[\rho_1,\rho_2]$ by
    \[
        d =  \max_{\rho \in [\rho_1,\rho_2]} \left\{ \frac{ \log \frac{ N_\rho(D) }{N_{\rho_1}(D)}}{ \log \frac{\rho_1}{\rho}  } \right\}
    \]
\end{definition}
Intuitively, this means that when we double the radii, the metric entropy, or number of covering spheres needed, decreases by a multiplicative factor of $2^d$.
This definition of fractal dimension is classical, but the reader will note also different in formulation from the intuitive one given in the main text.
However, the two notions are related given the bounded density assumption we made, from which immediately follows a bound on the number of points within each cluster.
On average, when we double the radius of a sphere around a point, the number of points in the larger sphere is roughly the number of points in the smaller sphere multiplied by $2^d$, because otherwise the spheres could not cover the space.
This latter behavior is what we measure when we talk about local fractal dimension around a point in the main paper.
However, because the number of points within each cluster must remain within a multiplicative constant of each other by the bounded density assumption, not only is this true on average, but the increase in number of points within a cluster has to be roughly uniform across all clusters.
Thus, this scaling behavior of points in a doubled sphere must be true everywhere in the data set, and thus we can connect the global average local fractal dimension alluded to in the main paper and measured for our data sets of interest to this more classical notion of fractal dimension based on covering spheres.

Recall that $k$ entries are selected as cluster centers for partitioning the database to result in clusters with maximum radius $r_c$.
From the definition above, when setting $\rho = r_c$, it is trivial to verify $ k \le N_{r_c} (D)$.
This upper bound is guaranteed by our requirement that the cluster centers not be within distance $r_c$.

Given any query $q$, the coarse search over the cluster centers always requires $k$ comparisons.
Additionally, the fine search is over the set $F$, defined to be the union of clusters with centers within distance $r+r_c$ from $q$.
As the time-complexity of similarity search is just the total of the coarse and fine searches, this implies that the total search time is $O(k + |F|)$.

By the triangle inequality, $F \subset B_D(q,r+2r_c)$,
so we can bound $|F| \le |B_D(q,r+2r_c)|$.
Let the fractal dimension $D$ at the scale between $r_c$ and $2r_c + r$ be $d$.
Recall that the local fractal dimension determines how many more points we hit when we scale the radius of a sphere.
Also, we use the density bound to give us that 
\[
    \frac{1}{\gamma} \left|B_D(p,\rho)\right| \le \mathbb{E}_q \left|B_D(q,\rho)\right| \le \gamma  \left|B_D(p,\rho)\right|
\]
for any choice of point $p$ and radius $\rho$.
Then
\begin{align}
    \mathbb{E}_q \left[ \left|B_D(q, r+2r_c)\right| \right] &\le \gamma \left|B_D(p,r+2r_c)\right| \\
                                             &\le \gamma \left|B_D(p,r)\right|\left(\frac{r+2r_c}{r}\right)^d \\
                                             &\le \gamma^2 \mathbb{E}_q \left[ \left|B_D(q,r)\right| \right] \left(\frac{r+2r_c}{r}\right)^d ,
\end{align}
because we are measuring the relative number of points found in spheres of radius $r$ vs radius $r+2r_c$ respectively for some particular point $p$.
But $\gamma$ is a constant, and disappears in asymptotic notation. Thus, total search time is in expectation over points $q$
\[
    O\left(k + \left|B_D(q,r)\right|\left(\frac{r+2r_c}{r}\right)^d \right).
\]
However, note that $k$ is linear in metric entropy and $|B_D(q,r)|$ is the output size, so similarity search can be performed in time linear to metric entropy and a polynomial factor of output size.
Provided that the fractal dimension $d$ is small and $k$ is large, the search time will be dominated by the metric entropy component, which turns out to be the regime of greatest interest for us.
We have thus proven bounds for the time-complexity of similarity search, given a self-similarity condition on the density of the dataset.

\subsubsection{Space-complexity}
Here we relate the space-complexity of our entropy-scaling similarity search data structure to information-theoretic entropy.
Traditionally, information-theoretic entropy is a measure of the uncertainty of a distribution or random variable and is not well-defined for a finite database.
However, the notion of information-theoretic entropy is often used in data compression as a shorthand for the number of bits needed to encode the database, or a measure of the randomness of that database.
We use entropy in the former sense; precisely, we define the entropy of a database as the number of bits needed to encode that database, a standard practice in the field.
Thus, we consider entropy-compressed forms of the original database, such as that obtained by Prediction by Partial Matching (PPM), Lempel-Ziv compression (e.g. Gzip), or a Burrows-Wheeler Transform (as in Bzip2), and use their size as an estimate of the entropy $S_{orig}$ of the database.

For all commonly used compression techniques, decompression time is linear in the size of the uncompressed data.
Obviously, even with linear decompression, decompressing the entire database for each similarity search would squander the entropy-scaling benefits of our
approach.
However, note that the fine search detailed above only needs access to a subset of clusters and furthermore needs full access to that set of clusters.
It is therefore always asymptotically `free' to decompress an entire cluster at once, if any member of that cluster needs to accessed.
Thus, one ready solution is to simply store entropy-compressed forms of each cluster separately.

Compressing each cluster separately preserves runtime bounds, but makes it difficult to compare the compressed clustered database size to the original compressed database size.
This results from the possibility that redundancy across clusters that would originally have been exploited by the compressor can no longer be exploited 
once the database is partitioned.
Intuitively, for any fixed-window or block compressor, grouping together similar items into clusters should increase the performance of the compressor, but it is unclear \textit{a priori} if that balances out the loss of redundancy across clusters.

A somewhat more sophisticated solution is to reorder the entries of the database by cluster, compress the entire database, and then store indexes into the starting offset of each cluster.
For popular tools such as Gzip or Bzip2, this is possible with constant overhead $\kappa$ per index.
Because the entire database is still being compressed, redundancy across clusters can be exploited to reduce compressed size, while still taking advantage of similar items being grouped together.
Thus, in expectation over uniformly-randomly chosen orderings of the database entries (obviously, there is some optimal ordering, but computing that is computationally infeasible), the compressed clustered database size $S_{clust} \le S_{orig}$.
Then, total expected space-complexity of our data structure is $O(\kappa k + S_{orig})$; recall here that $k$ is the number of clusters and is bounded by the metric entropy of the database.
Thus, space complexity is linear in metric entropy plus information-theoretic entropy.

Additionally, given that our distance function measures marginal information-theoretic entropies, we can also give a bound on the total information-theoretic entropy of the database by using metric entropy and the cluster radius.
Let $l$ be the maximum distance of two points in the space.
The na\"ive upper bound on total entropy is then $O(nl)$, where $n$ is the total number of points in the database, because distance and entropy are related.
Recall that we chose $k$ points as cluster centers, where $k$ is bounded by metric entropy, for a maximum cluster radius $r_c$.
Encoding each non-center point $p$ as a function of the nearest cluster center requires $O(n  r_c)$ bits.
Specifying the privileged points again requires $O(kl)$ bits, so together the total information-theoretic entropy is $O(kl + n r_c)$.
In other words, not only is space complexity linear in metric entropy plus information-theoretic entropy, but information-theoretic entropy itself is also bounded by the low-dimensional coarse structure of the database.

\subsubsection{Clustering time complexity}
Although clustering the database is a one-time cost that can further be amortized over future queries, we still require that cluster generation be tractable.
Here we present a trivial
$O(kn)$ algorithm for cluster generation with clusters of maximum radius $r_c$ that appears to work sufficiently well in practice (and is the algorithm used in esFragBag):
\begin{itemize}
    \item Initialize an empty set of cluster centers $C$. Let $\delta(x,C)$ be the distance from a point $x$ to $C$, defined to be $\infty$ if $C = \emptyset$.
    \item Randomly order the $n$ entries of the database $D = \{d_1, \ldots, d_n  \}$
    \item For $i = 1, \ldots n$,
        \begin{itemize}
            \item If $\delta(d_i,C) > r_c$, append $d_i$ to $C$.
        \end{itemize}
    \item For $i = 1, \ldots n$,
        \begin{itemize}
            \item Assign $d_i$ to the cluster represented by the nearest item in $C$.
        \end{itemize}
\end{itemize}
Because we need to compare each of $n$ items against up to $k$ items in $C$ in each of the for loops, this trivial algorithm takes $O(kn)$ time.

Additionally, insertions can be performed in $O(k)$ time.
For a new entry $d_{n+1}$, if $\delta(d_{n+1},C) \le r_c$, assign $d_{n+1}$ to the cluster represented by the nearest item in $C$.
Otherwise, append $d_{n+1}$ to $C$ as a new cluster center.
This clearly requires exactly $k$ comparisons to do.
Note that with insertions of this kind, items are no longer guaranteed to be assigned to the nearest cluster center; however, they are still guaranteed to be assigned to some cluster center within distance $r_c$, which is all that is needed for entropy-scaling to work.

Deletions are slightly more complicated.
If the entry to be deleted is not a cluster center, then removing it takes constant time.
However, if it is a cluster center, we effectively have to remove the entire cluster and reinsert all the non-center elements, which will take $O(k \cdot \textrm{[size of cluster]})$.
Thus, in expectation over a uniform random choice of item to be deleted, deletions can also be performed in $O(k)$ time.

\subsection{Ammolite}

\subsubsection{Simplification and compression}

Given a molecular graph, any vertex or edge that is not part of a simple cycle or a tree is removed, and any edge that is part
of a tree is removed.
This preserves the node count, but not the topology, of tree-like structures, and preserves simple cycles,
which represent rings in chemical compounds.
For example, as shown in Figure~\ref{fig:ammolite}, both caffeine and adenine would be reduced to a purine-like graph.

After this transformation is applied to each molecule in a database to be compressed, we identify all clusters
of fully-isomorphic transformed molecular graphs.
Isomorphism detection is performed using the VF2~\cite{cordella2001improved} 
algorithm; a simple hash computed from the
number of vertices and edges in each transformed molecular graph is first used 
to filter molecular graphs that cannot possibly be isomorphic.
A representative from each such cluster is stored in SDF format; collectively, these representatives form a 
``coarse'' database.
Along with each representative, we preserve the information necessary to reconstruct each original molecule,
as a pointer to a set of vertices and edges that have been removed or unlabeled.

Ammolite is implemented in Java, and its source code is available on Github.

\subsection{MICA}

\subsubsection{Alphabet Reduction}

Alphabet reduction---reducing the 20-letter standard amino acid alphabet to a
smaller set, in order to accelerate search or improve homology detection---has
been proposed and implemented several times~\cite{bacardit2007automated, peterson2009reduced}.
In particular, \citet{murphy2000simplified} considered reducing the
amino-acid alphabet to 17, 10, or even 4 letters.
More recently, \citet{zhao2012rapsearch2} and \citet{huson2013poor} applied a reduction to
a 4-letter alphabet, termed a ``pseudoDNA'' alphabet, in sequence alignment.

When using BLASTX for coarse search (which we call caBLASTX), we extend the 
compression approach of 
\citet{daniels2013compressive} using a reversible alphabet reduction.
We use the alphabet reduction of \citet{murphy2000simplified} to map the 
standard amino
acid alphabet (along with the four common ambiguous letters ) onto a 4-letter 
alphabet.
Specifically, we map F, W, and Y into one cluster; C, I, L, M, V, and J into
a second cluster, A, G, P, S, and T into a third cluster, and
D, E, N, Q, K, R, H, B, and Z into a fourth cluster.
By storing the offset of the original letter within each cluster, the original
sequence can be reconstructed, making this a reversible reduction.
This alphabet reduction is not used when using DIAMOND for coarse search, as
DIAMOND already relies on its own alphabet reduction.

\subsubsection{Database Compression}

Given a protein sequence database to be compressed, we proceed as follows:
\begin{enumerate}
        \item First, initialize a table of all possible $k$-mer seeds of
        our (possibly 4-letter reduced) alphabet, as well as a coarse database 
        of sequences, initially containing the (possibly reduced-alphabet)
        first sequence in the input database.
        \item For each $k$-mer of the first sequence, store its position in the
        corresponding entry in the seed table.
        \item For each subsequent sequence $s$ in the input, reduce its 
        alphabet and slide a window of 
        length $k$ along the sequence, skipping single-letter repeats of length
        greater than 10.
        \item
        \begin{enumerate}
        \item Look up these $k$ residues in the seed table.
        For every entry matching to that $k$-mer in the seed table, follow
        it to a corresponding subsequence in the coarse database and attempt
        \textit{extension} (defined below).
        If no subsequences from this window can be extended, move the window
        by $m$ positions, where $m$ defaults to 20.
        \item If a match was found via extension, move the $k$-mer window to
        the first $k$-mer in $s$ after the match, and repeat the extension
        process.
        \end{enumerate}
\end{enumerate}
        
Given a $k$-mer in common between sequence $s$ and a subsequence $s'$ pointed to by the
seed table, first attempt \textit{ungapped} extension:
\begin{enumerate}
        \item Within each window of length $m$ beginning with a $k$-mer match, 
        if there are at least 60\% matches between $s$ and $s'$, then there is 
        an ungapped match.
        \item Continue ungapped matching using $m$-mer windows until no more
        $m$-mers of at least 60\% sequence identity are found.
        \item The result of ungapped extension is that there is an alignment 
        between $s$ and $s'$ where the only differences are substitutions,
        at least 60\% of the positions contain exact matches.
\end{enumerate}
        
When ungapped extension terminates, attempt \textit{gapped} extension.
From the end of the aligned regions thus far, align 25-mer windows of both
$s$ and $s'$ using the Needleman-Wunsch~\cite{needleman1970general} algorithm 
using an identity matrix.
Note that the original caBLASTP~\cite{daniels2013compressive} used BLOSUM62 as 
it was
operating in amino acid space; as we are now operating in a reduced-alphabet
space, an identity matrix is appropriate, just as it is for nucleotide space.
After gapped extension on a window length of 25, attempt ungapped extension
again.

If neither gapped nor ungapped extension can continue, end the extension phase.
If the resulting alignment has less than 70\% sequence identity (in the 
reduced-alphabet space), or is shorter than 40 residues, discard it, and 
attempt extension on the next entry in the seed table for the original $k$-mer,
continuing on to the next $k$-mer if there are no more entries.

If the resulting alignment does have at least 70\% sequence identity in the
reduced-alphabet space, and is at least 40 residues long, then create a link
from the entry for $s'$ in the coarse database to the subsequence of $s$
corresponding to the alignment.
If there are unaligned ends of $s$ shorter than 30 residues, append them to the
match.
Longer unaligned ends that did not match any subsequences reachable from the
seed table are added into the coarse database themselves, following the same
$k$-mer indexing procedure as the first sequence.

Finally, in order to be able to recover the original sequence with its original
amino acid identities, a \textit{difference script} is associated with each
link.
This difference script is a representation of the insertions, deletions, and
substitutions resulting from the Needleman-Wunsch alignment, along with (if 
alphabet reduction is used) the
offset in each reduced-alphabet cluster needed to recover the original alphabet.
Thus, for example, a valine (V) is in the cluster containing C, I, L, M, V, and 
J.
Since it is the 4th entry in that 5-entry cluster, we can represent it with
the offset 4.
Since the largest cluster contains 9 elements, only four bits are needed to
store one entry in the difference script.
More balanced clusters would have allowed 3-bit storage, but at the expense of
clusters that less faithfully represented the BLOSUM62 matrix and the
physicochemical properties of the constituent amino acids.

Because of the seed table, compression is memory-intensive and CPU-intensive.
Compressing the September, 2014 NCBI NR database required approximately 39 hours on a 12-core Xeon with 128GB RAM.

\subsubsection{Query Clustering}

Metagenomic reads are themselves nucleotide sequences, so no alphabet reduction
is performed on them directly.
When BLASTX is used for coarse search (caBLASTX), MICA relies on query-side clustering.
Metagenomic reads are compressed using the same approach as the
protein database, without the alphabet reduction step and with a number of
different parameters.
The difference scripts for metagenomic reads do not rely on the cluster offsets,
but simply store the substituted nucleotides.

Furthermore, unlike protein databases, where most typical sequences range in 
length from 100 to over 1000 amino acids, next-generation sequencing reads are 
typically short and usually of fixed length, which is known in advance.
Thus, the minimum alignment length required for a match, and the maximum
length unaligned fragment to append to a match, require different values based
on the read length.

An additional complication is that insertions and deletions from one read to
another will change the reading frame, potentially resulting in 
different amino acid sequences.
For this reason, query clustering requires long, \emph{ungapped} windows of high
sequence identity.
Specifically, for 202-nucleotide reads, for two sequences to cluster together,
we require a 150-nucleotide ungapped region of at least 80\% sequence identity.

We note that unlike the compression of the database, which can be amortized 
over future queries, the time spent clustering and compressing the queries 
cannot be amortized.
Thus, we would not refer to the query clustering as entropy-scaling, but it
still provides a constant speed-up.
For this reason, we include the time spent clustering and compressing queries in the search time for MICA.
When using DIAMOND for coarse search, MICA does not perform query-side 
clustering, and instead relies on DIAMOND's indexing of the queries.

\subsubsection{Search}

Given a compressed protein database and a compressed query read set, search
comprises two phases.
The first, \emph{coarse search}, considers only the coarse sequences---the
representatives---resulting from compression of the protein database and the
query set.
When BLASTX is used for the coarse search (caBLASTX), each coarse nucleotide 
read is transformed into 
each of the six possible amino acid sequences that could result from it (three 
reading frames for both the sequence and its reverse complement).
Then, each of these amino acid sequences is then reduced back to a four-letter
alphabet using the same mapping as for protein database compression.
For convenience, the four-letter alphabet is represented using the standard
nucleotide bases, though this has no particular biological significance.
This is done so that the coarse search can rely on BLASTN (nucleotide BLAST) to
search these sequences against the compressed protein database.

For each coarse query representative (identified using a a coarse E-value of 
1000, along with the BLASTN arguments \texttt{-task blastn-short -penalty -1}; 
these arguments are recommended by the NCBI BLAST+ manual when queries are short), 
the set of coarse hits is used to
reconstruct all corresponding sequences from the original database by following
links to original sequence matches and applying their difference scripts.
The resulting \emph{candidates} are thus original sequences from the protein
database, in their original amino acid alphabet.
The query representative is also used to reconstruct all corresponding sequences
from the original read set.
Thus, for each coarse query representative, there is now a subset of the
metagenomic read set (the reads represented by that coarse query) and also a
subset of the protein database (the candidates).

When DIAMOND is used for coarse search, instead of an E-value threshold, the
argument \texttt{--top 60} is used; this causes DIAMOND to return all coarse
hits whose score is within 60\% of the top-scoring hit.
Without this argument, DIAMOND defaults to returning at most 25 hits for each
query sequence, which would result in significant loss of recall.

The second phase, \emph{fine search}, uses standard DIAMOND or BLASTX to translate each
of these reads associated with a coarse query representative and search for
hits only in the subset of the database comprising the candidates.
This fine search phase relies on a user-specified E-value threshold (or other
user-specified parameters to DIAMOND or BLASTX) to filter hits.
To ensure that E-value calculation is correct, the call to BLASTX uses a 
corrected database
size which is the size of the original, uncompressed protein database.

\subsubsection{Benchmarking}

Although our primary result is the direct acceleration of DIAMOND using our
entropy-scaling data structures, we also compared MICA to 
RapSearch2~\cite{zhao2012rapsearch2} version 2.22 and the November 29, 2014 
version of DIAMOND~\cite{buchfink2014fast}.
All tests were performed on a 12-core Intel Xeon X5690 running at 3.47GHz with
88GB RAM and hyperthreading; 24 threads were allowed for all programs.
Diamond was run with the \texttt{--sensitive} option.
In all cases, an E-value threshold of \texttt{1e-7} was used.

For the raw-read dataset, we filtered out reads starting or ending with 10 or 
more no-calls ('N').

MICA is implemented in Go, and its source code is available on Github.

\subsection{esFragBag}
We took the existing FragBag method as a black box and by design did not do anything clever in esFragBag except apply the entropy-scaling similarity search data structure.
We used a Go language implementation of FragBag, written by Andrew Gallant.
Additionally, we removed the sorting-by-distance feature of Andrew Gallant's 
FragBag search implementation, which does not improve the all-matching results we 
were interested in here---it lowers $k$-nearest neighbor search memory 
requirements while dominating the running time of $\rho$-nearest neighbor, the 
problem at hand.
This was done for both the FragBag and the esFragBag benchmarks, to ensure comparability.
All code was written in Go, and is available on Github.

The entire 2014 Oct 31 version of the Protein Data Bank was downloaded and the 
database was composed of fragment frequency vectors generated from all of the 
relevant PDB files using the 400-11.json fragment list \cite{budowski2010fragbag}.
For this paper, we implemented the benchmarking in Go, and have provided the 
source code for the benchmarking routine on Github.
This allowed us to benchmark just the search time, excluding the time to load the database from disk.
Note that the prototype implementation of esFragBag available only supports the 
all $\rho$-nearest neighbor search query found in FragBag.

\bibliographystyle{model5-names-noitalic}
\bibliography{main}